\documentclass{aa}  
\usepackage{graphicx}
\usepackage{txfonts}
\begin{document}
   \title{VLT Spectropolarimetry of the fast expanding Type Ia
   SN~2006X\thanks{Based on 
   observations made with ESO Telescopes at
   Paranal Observatory under Program IDs 076.D-0177 and 076.D-0178.}}
  \subtitle{}
   \author{F. Patat\inst{1}
   \and
    D. Baade\inst{1}
   \and
    P. H\"oflich\inst{2}
    \and
    J. R. Maund\inst{3,4}
    \and
    L. Wang\inst{5,6}
    \and
    J.C. Wheeler\inst{3}
}

   \offprints{F. Patat}

   \institute{European Organization for Astronomical Research in the 
	Southern Hemisphere (ESO), K. Schwarzschild-str. 2,
              85748, Garching b. M\"unchen, Germany\\
              \email{fpatat@eso.org}
              \and
	      Department of Physics, Florida State University, Tallahassee,
	      Florida 32306-4350, USA             \and
              Department of Astronomy and McDonald Observatory, 
              The University of Texas at Austin, Austin, TX 78712, USA
              \and
              Department of Physics, Texas A\&M University, College Station, 
	      Texas 77843, USA
	      \and
              Dark Cosmology Centre, Niels Bohr Institute, 
              University of Copenhagen, Juliane Maries Vej 30, 
              2100 Copenhagen, Denmark
              \and
              Purple Mountain Observatory, 2 West Bejing Road, 
	      Nanjing 210008, China
             }

   \date{Received July, 2008; accepted September 15, 2009}

 
  \abstract
   {}
   {The main goal of this study is to probe the ejecta geometry and to get 
    otherwise unobtainable information about the explosion mechanism of 
    Type Ia Supernovae (SNe).
   }
   {Using VLT-FORS1 we performed optical spectropolarimetric observations
   of the Type Ia SN~2006X on 7 pre-maximum epochs (day $-$10 to day $-$1)
   and one post-maximum epoch (+39 days).}
   {The SN shows strong continuum interstellar polarization reaching about
    8\% at 4000\AA, characterized by a wavelength dependency that is 
    substantially different from that of the Milky-Way dust mixture.
    Several SN features, like Si~II 6355\AA\/ and the Ca~II IR
    triplet, present a marked evolution. The Ca~II near-IR triplet shows a 
    pronounced polarization  ($\sim$1.4\%) already on day $-$10 in 
    correspondence with a strong high-velocity feature (HVF). The Si~II 
    polarization peaks on day $-$6 at about 1.1\% and decreases to 0.8\% 
    on day $-$1. By day +39 no polarization signal is detected for the Si~II 
    line, while the Ca~II IR triplet shows a marked re-polarization at the 
    level of 1.2\%. As in the case of another strongly polarized SN (2004dt), 
    no polarization was detected across the O~I 7774\AA\/ absorption.   
 }
   {The fast-expanding SN~2006X lies on the 
    upper edge of the relation between peak polarization and decline rate, and 
    it confirms previous speculations about a correlation between
    degree of polarization, expansion velocity, and HVF strength.
    The polarization of Ca~II detected in our last epoch, the most advanced
    ever obtained for a Type Ia SN, coincides in velocity with the outer 
    boundary of the Ca synthesized during the explosion (15,000-17,000 
    km s$^{-1}$) in delayed-detonation models. This suggests a large scale 
    chemical inhomogeneity as produced by off-center detonations, a rather 
    small amount of mixing, or a combination of both effects.
    In contrast, the absence of polarization at the inner edge of  
    the Ca-rich layer (8000-10,000 km s$^{-1}$) implies a substantial 
    amount of mixing in these deeper regions.}


\authorrunning{F. Patat et al.}
\titlerunning{VLT spectropolarimetry of the Type Ia SN~2006X}

   \maketitle
%

\section{\label{sec:intro}Introduction}

\begin{figure}
\centering
\includegraphics[width=8cm]{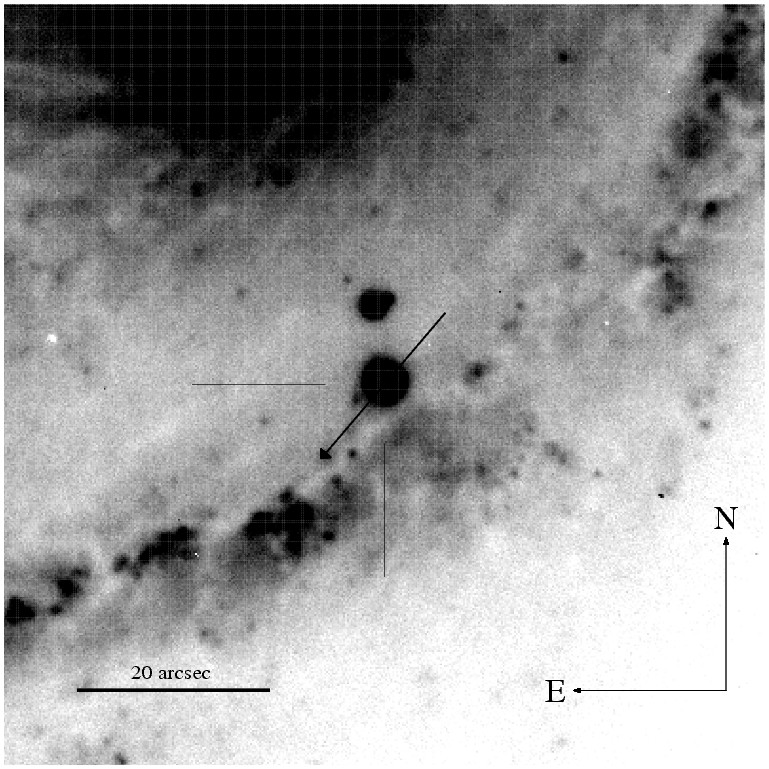}
\caption{\label{fig:fchart}SN~2006X in NGC~4321. The $R$ band image was 
obtained with VLT-FORS1 (see Sect.~\ref{sec:obs}) on February 11, 2006
(30 sec exposure time). The arrow placed at the SN position indicates
the inter-stellar polarization position angle (139.5 degrees; see
Sect.~\ref{sec:isp}).}
\end{figure}

The importance of Type Ia Supernovae (hereafter SNe~Ia) cannot be
overstated. Besides their fundamental role in the chemical enrichment
of galaxies, in the past ten years they have played a crucial part in
Cosmology, bringing new and exciting insights into the history of the
Universe and its future evolution (Riess et al. \cite{riess};
Perlmutter et al. \cite{perlmutter}).  For these reasons, SNe~Ia are
being studied by several extended surveys, which aim at pinning down
the cosmological parameters. Notwithstanding these extensive efforts,
the understanding of fundamental issues like the progenitor's nature,
the explosion mechanisms, and empirical relations such as luminosity
vs. light curve shape, necessarily rely on the study of the nearby
objects.

During the last decade, especially after the advent of 8m-class
telescopes, a new aspect of the SN phenomenon has come within the
reach of investigation, i.e. the geometry of the explosion. Since the
electron scattering in asymmetric SN ejecta is expected to produce a
net linear polarization (see for instance H\"oflich
\cite{hoeflich}), the ejecta geometry and clumping can be directly 
probed by means of polarimetry. Starting with the first broad-band
studies (Wang et al. \cite{wang96}), two groups have been collecting
unique information through the spectropolarimetric study of a number
of SNe~Ia (Wang et al. \cite{wang97}, \cite{wang01}; Howell et
al. \cite{howell}; Kasen et al. \cite{kasen}; Wang et
al. \cite{wang03}; Hoeflich et al. \cite{hoeflich06}, Leonard et
al. \cite{leonard}; Wang et al. \cite{wang06b}, Chornock \& Filippenko
\cite{chornock}) that have led to important insights on explosion
asymmetry and ejecta clumping that no other technique can provide
(Leonard et al. \cite{leonard}; Wang, Baade \& Patat \cite{wang07};
Wang \& Wheeler \cite{WW08}).

Here we present low-resolution optical spectropolarimetry of SN~2006X
obtained at the ESO Very Large Telescope, spanning from 10 days before
$B$ maximum light to about 40 days past maximum. To the best of our
knowledge, in terms of evolution coverage, this is the most complete
spectropolarimetric data set ever obtained for a Type Ia SN, equaled
only by SN~2001el (Wang et al. \cite{wang03}).

SN~2006X was discovered by Suzuki \& Migliardi (\cite{suzuki}) on 4
February 2006 (UT) in the Virgo Cluster spiral galaxy NGC~4321 (see
Fig.~\ref{fig:fchart}), located at a distance of 16.1$\pm$1.3 Mpc
(Ferrarese et al. \cite{ferrarese}) and receding with a velocity
v$_{gal}$=1572 km s$^{-1}$ (Rand \cite{rand}). A few days later, the
object was classified as a normal Type Ia event 1-2 weeks before
maximum light and most likely affected by a substantial extinction
(Quimby, Brown \& Gerardy \cite{quimby}). Prompt VLA observations have
shown no radio source at the SN position (Stockdale et
al. \cite{stockdale}), actually establishing one of the deepest and
earliest limits for radio emission from a Type Ia, implying a
mass-loss rate of less than a few 10$^{-8}$ M$_\odot$ yr$^{-1}$, for a
wind velocity of 10 km s$^{-1}$ and assuming a spherically symmetric,
steady wind. Observations carried out with XRT on board Swift put an
upper limit of 1.5$\times$10$^{39}$ erg s$^{-1}$ for the X-ray flux in
the 0.2-10 keV band (Immler et al. \cite{immler}), corresponding to a
mass loss rate lower than 7$\times$10$^{-6}$ M$_\odot$
yr$^{-1}$. The SN was also reported to show a rather strong continuum
linear polarization in the optical domain, reaching 8 per cent in the
blue and decreasing to 3.5 per cent in the red, with a wavelength
dependency significantly different from that typical of extinguished
stars in our own galaxy (Wang et al. \cite{wang06a}). High resolution
spectroscopy around maximum light has shown very intense interstellar
absorption features, identified as CN, Na~I, Ca~II and K~I, confirming
the presence of significant extinction along the line of sight
(Lauroesch et al. \cite{lauroesch}; Patat et al. \cite{patat07}; Cox
\& Patat \cite{cox}). Rather remarkably, the multi-epoch high
resolution spectroscopy has also revealed a pronounced evolution in a
number of the Na~ID components, which have been interpreted as
signatures of circum-stellar material lost by the progenitor system
before the explosion (Patat et al. \cite{patat07}. But see also Chugai
\cite{chugai}). An extensive optical
and near-infrared study has been presented by Wang et
al. (\cite{wangx}). They have shown that this SN has one of the
highest expansion velocities ever observed, and this is accompanied by
slightly peculiar light and color curves. These might be explained by
a light echo and/or the interaction between the SN ejecta and the
circumstellar material (see also Wang et al. \cite{wangx08}; Crotts \&
Yourdon \cite{crotts}).  The reddening corrected value for the decline
rate of SN~2006X is $\Delta m_{15}(B)=$1.31$\pm$0.05. The
normalized ($\Delta m_{15}$=1.1) absolute $V$ magnitude is
$-$19.2$\pm$0.2, which is consistent with those of fiducial Type Ia
SNe, while it is 0.2-0.3 mag fainter in $I$, $J$ and $K$ bands (Wang
et al. \cite{wangx}).

In this paper we report the results of a spectropolarimetric campaign
conducted with VLT-FORS1, covering seven pre-maximum phases from day
$-$10 to day $-1$. The data set includes an additional epoch 39 days
past maximum light. The paper is organized as follows. In
Sect.~\ref{sec:obs} we present and discuss our observations and data
reduction. The spectroscopic evolution during the phases covered by
our observations is analyzed in
Sect.~\ref{sec:evol}. Sect.~\ref{sec:isp} deals with inter-stellar
polarization, while Sect.~\ref{sec:snpol} presents the SN intrinsic
polarization.  Finally, in Sect.~\ref{sec:disc} we discuss our results
and summarize our conclusions. Additional figures are presented in
Appendix A, available online.

\begin{figure}
\centering
\includegraphics[width=8cm]{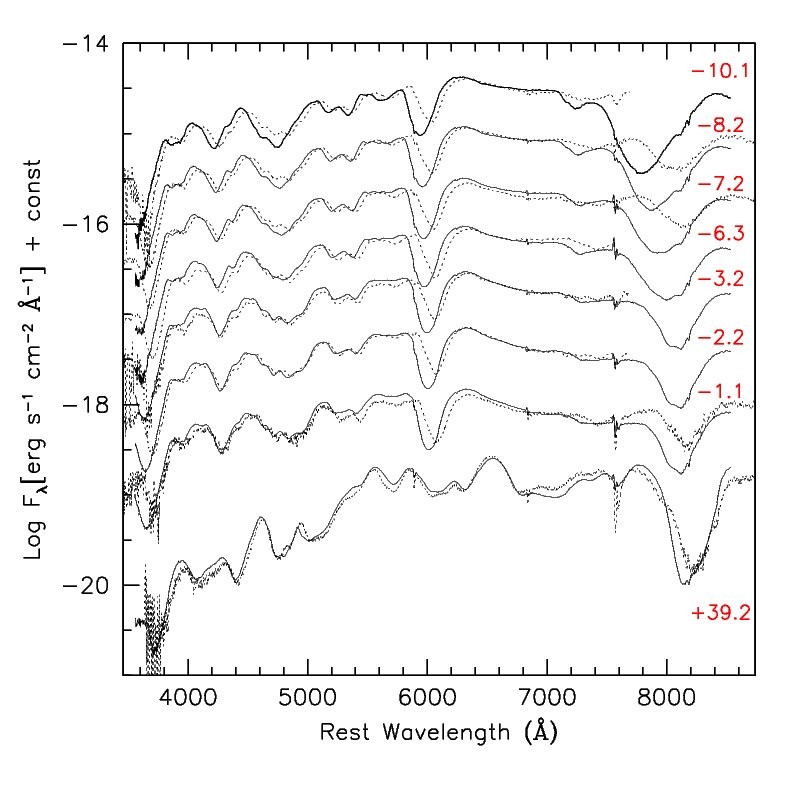}
\caption{\label{fig:evol}Spectral evolution of SN~2006X. Phases refer to 
$B$ maximum light (JD=2,453,786.17; Wang et al. \cite{wangx}). For
comparison, the spectra of SN~2002bo (Benetti et al. \cite{benetti04})
are also plotted (dotted lines). For presentation, the observed
spectra of SN~2002bo (not corrected for extinction) were
reddened using a standard extinction law ($R_V$=3.1) and
$E_{B-V}$=0.8.}
\end{figure}

\section{\label{sec:obs}Observations and Data Reduction}

\begin{table}
\tabcolsep 1.2mm
\caption{\label{tab:obs}Log of VLT-FORS1 spectropolarimetric observations 
of SN~2006X.}
\begin{tabular}{cccccc}
\hline
Date  & MJD                & phase       & $V$ & airmass     & exp. time\\
(UT)  & (JD-2,400,000.5)   & (days)      & (*) &  (average)  & (seconds) \\
\hline
2006-02-09 & 53775.37 & $-$10.1 & 15.3&1.36  & 4$\times$300\\
2006-02-11 & 53777.29 & $-$8.2  & 14.9&1.41  & 2$\times$4$\times$300\\
2006-02-12 & 53778.29 & $-$7.2  & 14.7&1.50  & 2$\times$4$\times$600\\
2006-02-13 & 53779.22 & $-$6.3  & 14.5&1.68  & 4$\times$600\\
2006-02-16 & 53782.32 & $-$3.2  & 14.3&1.43  & 4$\times$600\\
           &          &         &     &      & 4$\times$420\\
2006-02-17 & 53783.31 & $-$2.2  & 14.2&1.39  & 2$\times$4$\times$600\\
2006-02-18 & 53784.39 & $-$1.1  & 14.2&1.57  & 4$\times$480\\
2006-03-30 & 53824.20 & +38.7   & 15.9&1.34  & 2$\times$4$\times$900\\
2006-03-31 & 53825.20 & +39.7   & 15.9&1.34  & 2$\times$4$\times$900\\
\hline
\multicolumn{6}{l}{(*) $V$ magnitudes are derived from Wang et al. 
(\cite{wangx}).}
\end{tabular}
\end{table}

We observed SN~2006X on 9 different epochs, using the FOcal
Reducer/low-dispersion Spectrograph (hereafter FORS1), mounted at the
Cassegrain focus of the ESO--Kueyen 8.2m telescope (Appenzeller et al.
\cite{appenzeller}). In this multi-mode instrument, equipped with a
2048$\times$2048 pixel (px) TK2048EB4-1 backside-thinned CCD,
polarimetry is performed introducing a Wollaston prism
(19$^{\prime\prime}$ throw) and a super-achromatic half-wave plate. In
order to reduce some known instrumental problems (see Patat \&
Romaniello \cite{patat06}) we always used 4 half-wave plate (HWP)
angles (0, 22.5, 45 and 67.5 degrees). Exposure times ranged from 5 to
15 minutes per plate angle and, in some cases, the sequence was
repeated in order to increase the signal-to-noise ratio. All spectra
were obtained with the low-resolution G300V grism coupled to a 1.1
arcsec slit, giving a spectral range 3300-8600 \AA, a dispersion of
$\sim$2.9 \AA\/ pixel$^{-1}$ and a resolution of 12.4 \AA\/ (FWHM) at
5800 \AA. Data were bias, flat-field corrected and wavelength
calibrated by means of standard tasks within IRAF\footnote{IRAF is
distributed by the National Optical Astronomy Observatories, which are
operated by the Association of Universities for Research in Astronomy,
under contract with the National Science Foundation.}. The RMS error
on the wavelength calibration is about 0.7
\AA. The ordinary (upper) and extraordinary (lower) beams were processed 
separately. Stokes parameters, linear polarization degree and position
angle were computed by means of specific routines written by
us. Finally, polarization bias correction and error estimates were
performed following the prescriptions described by Patat \& Romaniello
(\cite{patat06}), while the HWP zeropoint angle chromatism was
corrected using the tabulated data (Jehin, O'Brien \& Szeifert
\cite{fors}). In order to increase the signal-to-noise ratio, 
multiple data sets obtained at the same epoch were combined and the
final Stokes parameters binned in $\sim$26 \AA\/ wide bins (10 pixels).
The last two epochs (March 30 and 31) were combined into a single
epoch (+39.2).

Flux calibration was achieved through the observation of
spectrophotometric standard stars with the full polarimetric optics
inserted (HWP angle set to 0). Instrumental polarization and
position angle offset were checked by observing polarized and
unpolarized standard stars, obtained within the FORS1 calibration
plan. Finally, the wavelength scale was corrected to the
rest-frame using the host galaxy recession velocity (1572 km s$^{-1}$,
Rand \cite{rand}).

Because of the rather heavy extinction suffered by SN~2006X (Quimby et
al. \cite{quimby}; Lauroesch et al. \cite{lauroesch}; Wang et
al. \cite{wangx}; Elias-Rosa et al. \cite{nancy}), the signal-to-noise
ratio shortwards of 3900\AA\/ is quite poor. As a consequence, the
polarization is not reliably measurable at those short wavelengths.
This prevents us from studying the behavior of the Ca~II H\&K
lines that fall in this region. 

The log of observations is reported in Table~\ref{tab:obs}, where the SN
phases were computed with respect to the $B$ maximum light
(JD=2,453,786.17; Wang et al. \cite{wangx}).

\begin{figure}
\centering
\includegraphics[width=8cm]{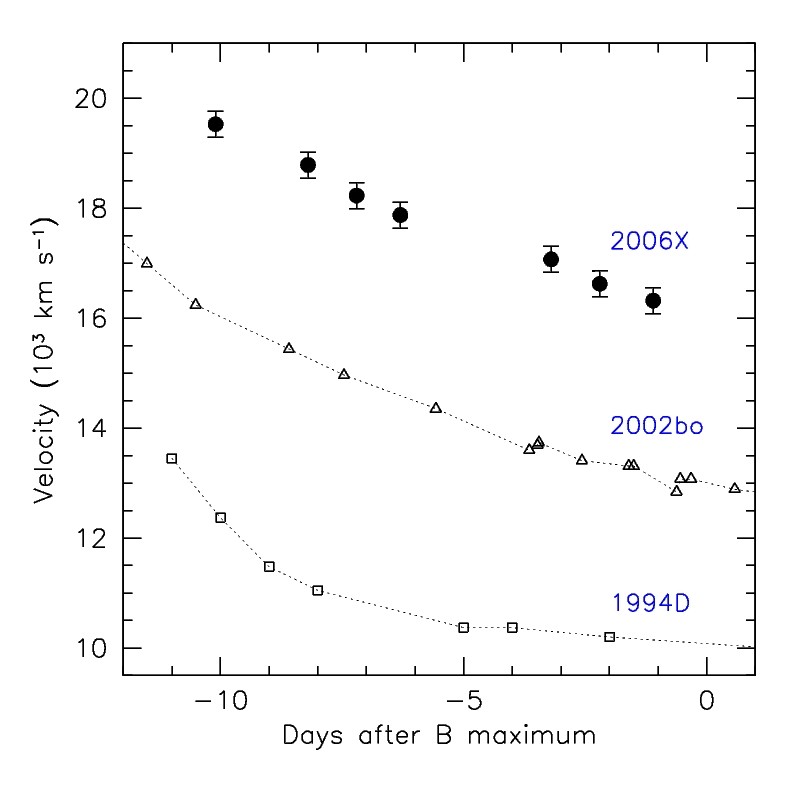}
\caption{\label{fig:vel}Evolution of the expansion velocity
deduced from the minima of the Si~II 6355\AA\/ absorption. For comparison,
the data of SN1994D (Patat et al. \cite{patat96}) and SN~2002bo (Benetti et al.
\cite{benetti04}) are also shown.}
\end{figure}

\section{\label{sec:evol}Spectroscopic Evolution}

\begin{figure}
\centering
\includegraphics[width=8cm]{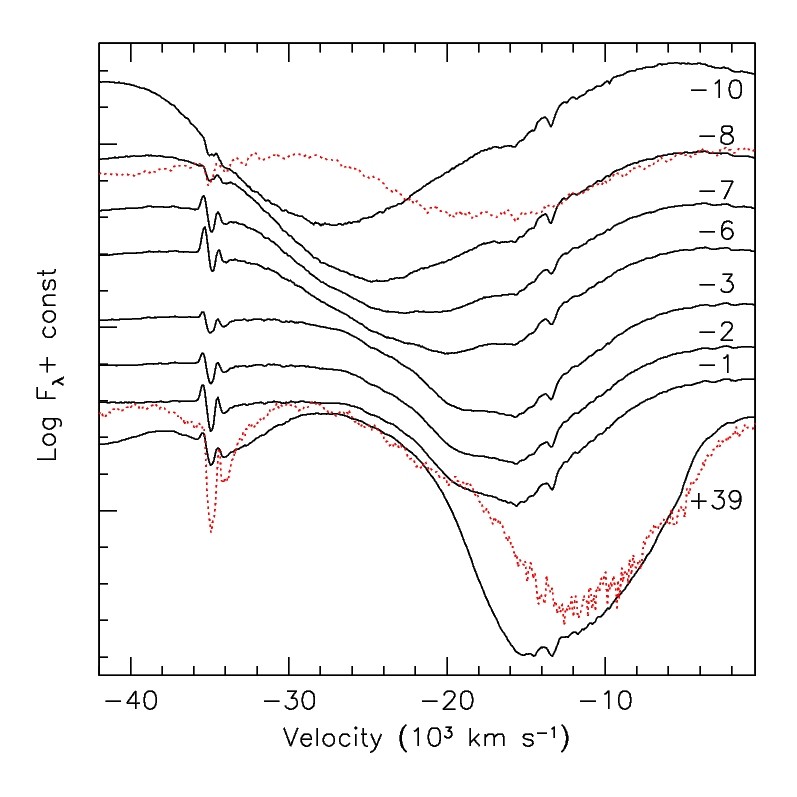}
\caption{\label{fig:ca}Evolution of the Ca~II IR triplet velocity profile. 
Velocities are relative to the $gf$-values weighted average triplet
wavelength (8567\AA). The dotted, light-colored curves are the spectra
of SN~2002bo (Benetti et al. \cite{benetti04}) at 8.5 days before
maximum (top) and 39.9 days after maximum light (bottom).}
\end{figure}

The maximum/post-maximum spectroscopic evolution of SN~2006X has been
presented by Wang et al. (\cite{wangx}), to which we refer the reader
for a detailed discussion. Their phase coverage extends up to about
300 days after maximum light and hence our data set is complementary,
in the sense that ours mostly covers the pre-maximum epochs.  The
spectroscopic evolution during these early phases is presented in
Fig.~\ref{fig:evol} where SN~2006X ($\Delta
m_{15}(B)=$1.31$\pm$0.05, $M_V$=$-$19.2$\pm$0.2; Wang et
al. \cite{wangx}) is compared to SN~2002bo ($\Delta
m_{15}(B)=$1.13$\pm$0.05, $M_V$=$-$19.42$\pm$0.33; Benetti et
al. \cite{benetti04}), which was also significantly polarized (Wang et
al. \cite{wang07}. See also Fig.~\ref{fig:sipol} here). The two
objects are indeed similar (see Wang et al. \cite{wangx} for a
thorough comparison), even though SN~2006X shows definitely higher
expansion velocities, especially at early epochs. This difference
becomes milder and milder as time goes by and at about 40 days past
maximum light the two objects are almost indistinguishable. At later
epochs the largest discrepancy is seen in the near-IR Ca~II triplet
absorption trough which, in SN~2006X, extends much further into the
blue (see the discussion below and Fig.~\ref{fig:ca}). The evolution
of the expansion velocities deduced from the minima of
the Si~II 6355\AA\/ absorption is presented in Fig.~\ref{fig:vel}
where, for comparison, we have included the data of two other normal
SNIa: 1994D (Patat et al. \cite{patat96}) and 2002bo (Benetti et
al. \cite{benetti04}).

As already pointed out by Wang et al. (\cite{wangx}), the expansion
velocities deduced for SN~2006X are exceptionally high (see their
Fig.~16 and Fig.~\ref{fig:vel} here), reaching $\sim$19,500 km
s$^{-1}$ 10 days before $B$ maximum light, i.e.  about 3500 km
s$^{-1}$ higher than in SN~2002bo which, from this point of view, was
already judged as an unusual event (Benetti et al. \cite{benetti04},
\cite{benetti05}). SN~2006X shows expansion
velocities which are even higher than in the extreme case of SN~1984A
(Benetti \cite{benetti89}), placing SN~2006X at the upper end of the
observed velocity range. During the phases covered by our
observations, the velocity decrease is very smooth ($\dot{\rm
v}\simeq$$-$350 km s$^{-1}$ day$^{-1}$), as in the case of SN~2002bo,
at variance with what is seen in objects like SN~1994D (see
Fig.~\ref{fig:vel}) that show significantly lower velocities and a
break around 5 days before maximum light (Patat et al. \cite{patat96};
Benetti et al. \cite{benetti04}). Clearly, as concluded by Wang et
al. (\cite{wangx}), SN~2006X belongs to the High Velocity Gradient
(HVG) class proposed by Benetti et al. (\cite{benetti05}), that
roughly coincides with the broad-line class of Branch et
al. (\cite{branch06}).

The value of the $\cal R$(SiII) parameter (Nugent et
al. \cite{nugent95}) for SN~2006X, estimated from the day $-$1 spectrum,
is $\cal R$(SiII)=0.07$\pm$0.04, which is exceptionally low if
compared to other HVG objects like SN~2002bo (0.17$\pm$0.05, Benetti
et al. \cite{benetti04}) at maximum light. For comparison, Wang et
al. (\cite{wangx}) report $\cal R$(SiII)=0.12$\pm$0.06.  A similarly
low value, 0.05$\pm$0.02, is obtained for the equivalent width ratio
of the two Si~II lines, as in Hachinger, Mazzali \& Benetti
(\cite{hachinger}). Nevertheless, given the fact that the features in
SN~2006X are very broad, it is not clear whether this low value is
really intrinsic or is rather due to the blending of the red wing of
the Si~II 5972\AA\/ line with the blue wing of the Si~II 6355\AA\/
line (see also Fig.~\ref{fig:evol}).

\begin{figure}
\centering
\includegraphics[width=8cm]{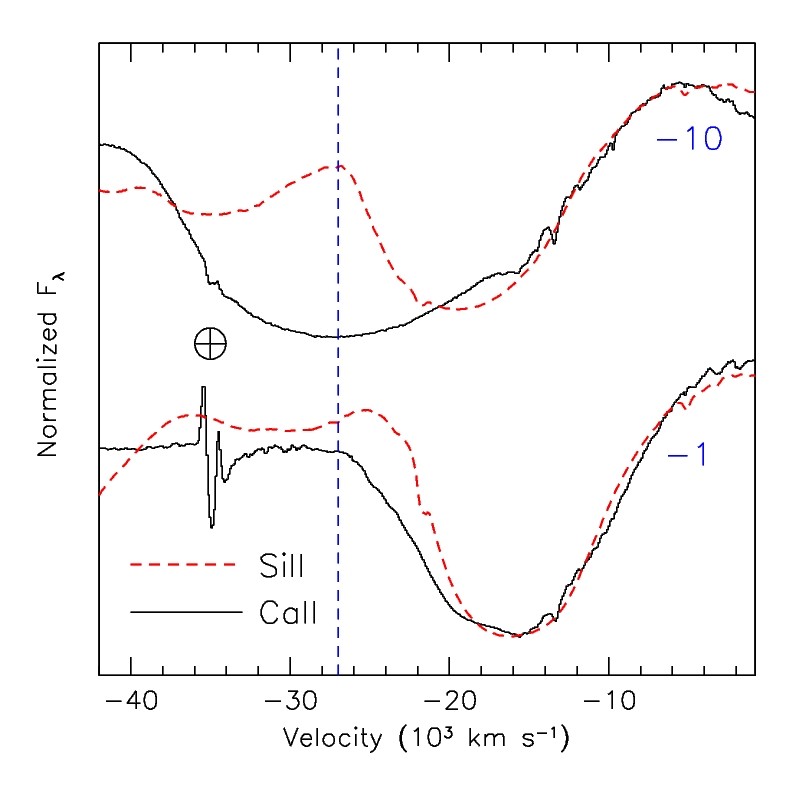}
\caption{\label{fig:casi}Si~II and Ca~II line profiles on day $-$10 (top)
and $-$1 (bottom). The vertical dashed line marks the Si~II cutoff
velocity for day $-$10.}
\end{figure}

Another interesting aspect shown by SN~2006X is the presence of a high
velocity feature (hereafter HVF), clearly visible in the Ca~II IR
triplet (see Fig.~\ref{fig:ca} and the discussion in Wang et
al. \cite{wangx}). These features are common to almost all SNe~Ia,
although with different strength and duration (Wang et
al. \cite{wang03}; Mazzali et al. \cite{mazzali05a}). In the earliest
spectrum of SN~2006X (day $-$10), the photospheric component
($\sim$16,000 km s$^{-1}$) is accompanied by a much broader HVF at
about 28,000 km s$^{-1}$, with material expanding at terminal
velocities as high as $\sim$40,000 km s$^{-1}$, similar to that
seen in SN~2002dj (see Mazzali et al. \cite{mazzali05a}, their
Fig.~2).  As in the case of the Si~II 6355\AA\/ line, SN~2006X shows
much higher Ca~II velocities than SN~2002bo (Fig.~\ref{fig:ca}, dotted
curve), for which Mazzali et al. (\cite{mazzali05a}) have identified a
HVF at about 20,000 km s$^{-1}$.

While discussing the Type Ia SN~2005cg, Quimby et
al. (\cite{quimby06}) have pointed out a sharp cutoff in the blue wing
of the Si~II 6355\AA\/ line, which they interpreted as the truncation
in the velocity distribution generated when the outer ejecta are
decelerated by the interaction with CSM. They also noticed that the
cutoff velocity appears to align with the blue edge of the Ca~II high
velocity absorption minimum, supporting the model presented by Gerardy
et al. (\cite{gerardy}), who attribute these HVFs to a surrounding,
H-rich region swept up by the SN ejecta. Quimby et
al. (\cite{quimby06}) have also shown that this is a common behavior
among spectroscopically normal Ia.

A similar analysis indicates that this is possibly the case also for
SN~2006X, which however presents a smoother roll-over from the Si~II
line blue wing to the continuum than SN~2005cg (Fig.~\ref{fig:casi}).
Even though the blue wing of the Si~II line is probably disturbed by
an adjacent feature, ten days before maximum light it appears to
extend up to $-$27,000 km s$^{-1}$, which coincides with the blue side
of the minimum of the Ca~II absorption trough. Nine days later, the
two features have a similar profile (see Fig.~\ref{fig:casi}).

In conclusion, from a spectroscopic point of view SN~2006X appears as
a normal object, even though it has shown expansion velocities which
are amongst the highest ever observed for a Type Ia SN.

\section{\label{sec:isp}Interstellar Polarization and Extinction Law}

\begin{figure}
\centering
\includegraphics[width=8cm]{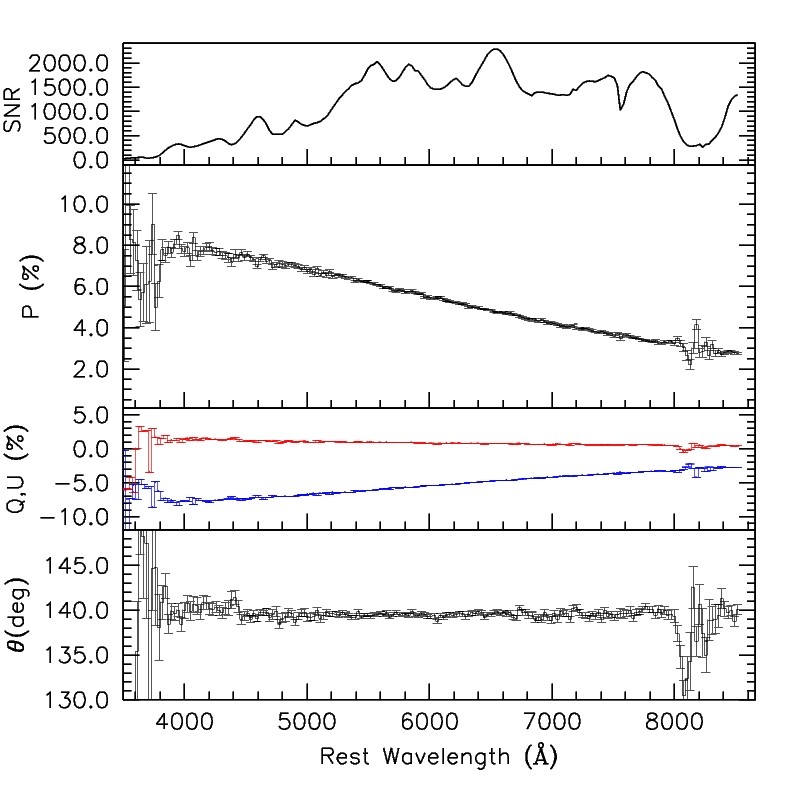}
\caption{\label{fig:isp} Spectropolarimetric data of SN~2006X for the 
last epoch (+39.2). From top to bottom: signal-to-noise ratio (SNR) on
the flux spectrum, linear polarization degree (P), Stokes parameters
(Q, U) and polarization angle ($\theta$). The errorbars indicate
1-sigma level statistical errors.}
\end{figure}

This SN is known to suffer from substantial extinction (Quimby et
al. \cite{quimby}, Lauroesch et al. \cite{lauroesch}; Wang et
al. \cite{wangx}) and, therefore, a significant interstellar
polarization (hereafter ISP) is expected (Serkowski, Matheson \& Ford
\cite{serkowski}).

The high resolution spectroscopy (Lauroesch et al. \cite{lauroesch};
Patat et al. \cite{patat07}; Cox \& Patat \cite{cox}) has revealed a
system of clouds within the host galaxy, and the presence of a
distinct molecular cloud with an exceptionally intense
$B^2\Sigma-X^2\Sigma$ CN vibrational band, accompanied by CH, CH$^+$,
Ca~I and diffuse interstellar bands. Stars having comparable features
in our own Galaxy are known to have color excesses larger than
$E_{B-V}$=1 (Crutcher \cite{crutcher}).  Additionally, a small
contribution to the extinction is produced by the Milky Way
($A_B\simeq$0.1; Schlegel, Finkbeiner \& Davis \cite{schlegel}).
Given the strong correlation between extinction and linear
polarization (Serkowski et al. (\cite{serkowski}), a significant
amount of ISP is expected for SN~2006X. This is indeed the case, as
clearly shown in Fig.~\ref{fig:isp}, where we have combined the data
sets of the last two epochs (+39 and +40 days) to increase the
signal-to-noise ratio. With the only remarkable exception being the
feature corresponding to the absorption minimum of the near-IR CaII
triplet (see next section), there is a rather strong continuum
polarization, which steadily decreases from $\sim$8\% at 4000\AA\/ to
$\sim$3\% at 8500\AA, as already reported by Wang et
al. (\cite{wang06b}). This wavelength dependence of the polarization
degree is rather different from what is seen in highly reddened stars
in our own Galaxy (Serkowski, Matheson \& Ford
\cite{serkowski}). Although it is possible to fit the data with the
empirical Serkowski:

\begin{displaymath}
P(\lambda) = P_{max} \; \exp \left (-K\; \ln^2 \frac{\lambda_{max}}{\lambda}
\right )
\end{displaymath}

\begin{figure}
\centering
\includegraphics[width=8cm]{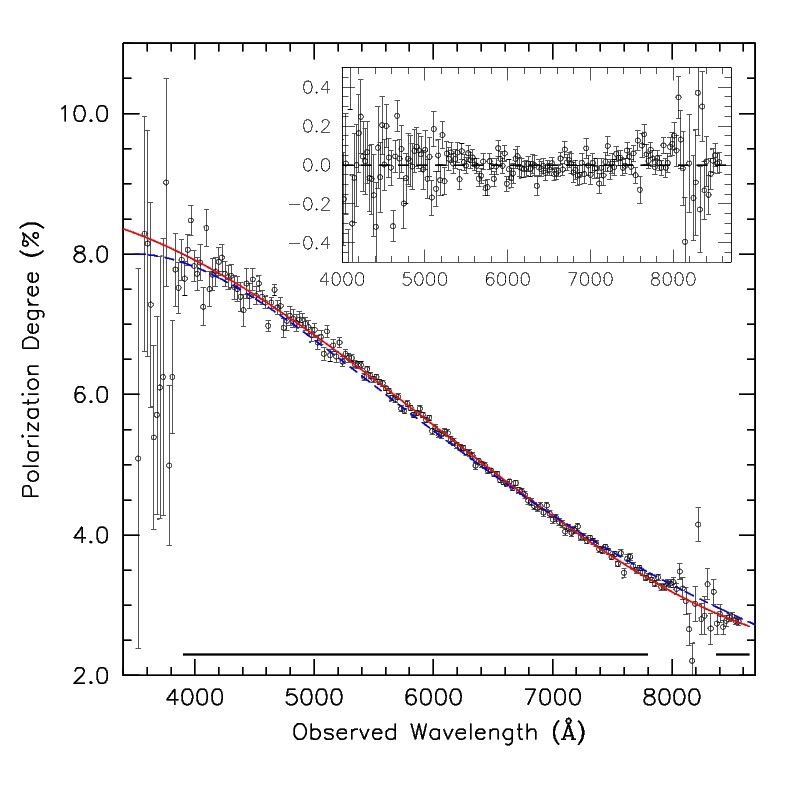}
\caption{\label{fig:ispdeg} The degree of linear polarization for the 
last epoch data (+39 d). The solid curve is a 3rd order polynomial fit to 
the observed data in the wavelength ranges indicated by the horizontal
segments close to the x-axis scale.  The dashed curve is a best fit
using a Serkowski law ($P_{max}$=8.0\%, $\lambda_{max}$=3500\AA\/ and
$K$=1.30). The upper right insert shows the residuals from the best
fit polynomial relation.}
\end{figure}

\noindent
the resulting values of $K$ (1.3$\pm$0.1) and $\lambda_{max}$
(3500$\pm$100\AA) are not compatible with the relation found by
Witthet et al. (\cite{whittet} for Galactic stars ($K = 0.01 + 1.66\;
\lambda_{max}$, where $\lambda_{max}$ is expressed in microns). In SN~2006X 
the wavelength dependency is much steeper (Fig.~\ref{fig:ispdeg}).  We
interpret this as an indication that the dust mixture is significantly
different from that typically found in the Milky Way (see also Wang et
al. \cite{wang06a}). This might well be related to the anomalous
strength of the CN absorption seen in high resolution spectroscopy
(Lauroesch et al. \cite{lauroesch}; Patat et al. \cite{patat07}),
indicating that the bulk of extinction occurred in a cold molecular
cloud (Patat et al. \cite{patat07}; Cox \& Patat
\cite{cox}).

Therefore, the ratio of total to selective extinction $R_V=A_V/E(B-V)$
derived for this SN must be significantly different from what is
commonly measured in the Galaxy. Serkowski et al. (\cite{serkowski})
have shown that the observed peak of polarization moves bluewards as
$R_V$ decreases and this is interpreted in terms of dust grain size
(see Whittet \cite{whittetb} and references therein). Therefore, the
ISP law observed in SN~2006X, characterized by a very blue peak
maximum, hints at a low value of $R_V$.  Interestingly, the
photometric and spectroscopic analysis by Wang et al. (\cite{wangx})
and Elias-Rosa et al. (\cite{nancy}) give $R_V$=1.48$\pm$0.06
($E(B-V)$=1.41$\pm$0.04) and $R_V$=1.7$\pm$0.1 ($E(B-V)$=1.2$\pm$0.3),
respectively. It is worth emphasizing that polarimetry gives an
independent proof that the nature of the intervening dust is different
from that of a typical Milky Way mixture. Interestingly, clear
deviances from the Galactic Serkowski law have been detected in other
SNe (Leonard \& Filippenko \cite{leonard01}; Leonard et al.
\cite{leonard02}; Maund et al. \cite{maund}) and in the bright optical 
transient in NGC~300 (Patat et al. \cite{patat09}).

\begin{table}
\caption{\label{tab:isp}ISP linear least squares fitting parameters 
for the wavelength range 6400-7200\AA.}
\tabcolsep 4.7mm
\begin{tabular}{ccccc}
\hline
Phase & $b$  & $\sigma$ & $P_{6800}$ \\
(days)& (\% $\AA^{-1}$) & (\%)     & (\%) \\    
\hline
$-$10 &	  $-$0.0013$\pm$0.00008 & 0.10 & 4.57$\pm$0.05 \\
$-$8  &	  $-$0.0014$\pm$0.00007 & 0.09 & 4.42$\pm$0.04 \\
$-$7  &	  $-$0.0014$\pm$0.00004 & 0.05 & 4.51$\pm$0.03 \\
$-$6  &	  $-$0.0013$\pm$0.00005 & 0.06 & 4.45$\pm$0.03 \\
$-$3  &	  $-$0.0013$\pm$0.00004 & 0.04 & 4.52$\pm$0.02 \\
$-$2  &	  $-$0.0013$\pm$0.00003 & 0.04 & 4.50$\pm$0.02 \\
$-$1  &	  $-$0.0012$\pm$0.00005 & 0.06 & 4.54$\pm$0.03 \\
$+$39 &	  $-$0.0012$\pm$0.00005 & 0.04 & 4.51$\pm$0.02 \\
\hline
\end{tabular}
\end{table}

Contrary to most polarimetric observations of SNe~Ia, for which it is
rather difficult to estimate the ISP contribution (see for instance
Leonard et al. \cite{leonard} or Wang et al. \cite{wang06b}), for
SN~2006X this is fairly simple, since the continuum polarization is
heavily dominated by the interstellar component.  Assuming that the
high continuum polarization is generated by intervening interstellar
material, the ISP Stokes parameters can be estimated as follows:
\begin{eqnarray}
\label{eq:isp}
Q_{ISP}(\lambda) & = & P_{ISP}(\lambda) \; \cos(2 \;\theta_{ISP}) \nonumber\\
U_{ISP}(\lambda) & = & P_{ISP}(\lambda) \; \sin(2 \;\theta_{ISP})
\end{eqnarray}

\noindent
Rather than adopting a Serkowski law for the degree of polarization,
we have preferred to use a 3rd order polynomial fit to the data,
performed in the wavelength ranges 3900-7800\AA\/ and 8350-8650\AA\/
in order to avoid low signal-to-noise data and to exclude the
Ca~II IR triplet region, where a polarization signal from the SN is
still present. Figure~\ref{fig:ispdeg} shows the best fit corresponding
to the following values for the polynomial coefficients:

\begin{equation}
\label{eq:ispdeg}
P_{ISP}(\lambda)=5.6-13.2(\lambda-\lambda_0)-0.1(\lambda-\lambda_0)^2+36.9(\lambda-\lambda_0)^3
\end{equation}

\noindent
where $P_{ISP}$ is in per cent, $\lambda$ is expressed in microns
and $\lambda_0$=0.6 $\mu$m. The RMS deviation from the best fit
relation is about 0.1\%, in line with the measurement accuracy (the
statistical RMS uncertainties on the polynomial coefficients are 0.02,
0.16, 0.68, and 4.76 respectively).

\begin{figure}
\centering
\includegraphics[width=8cm]{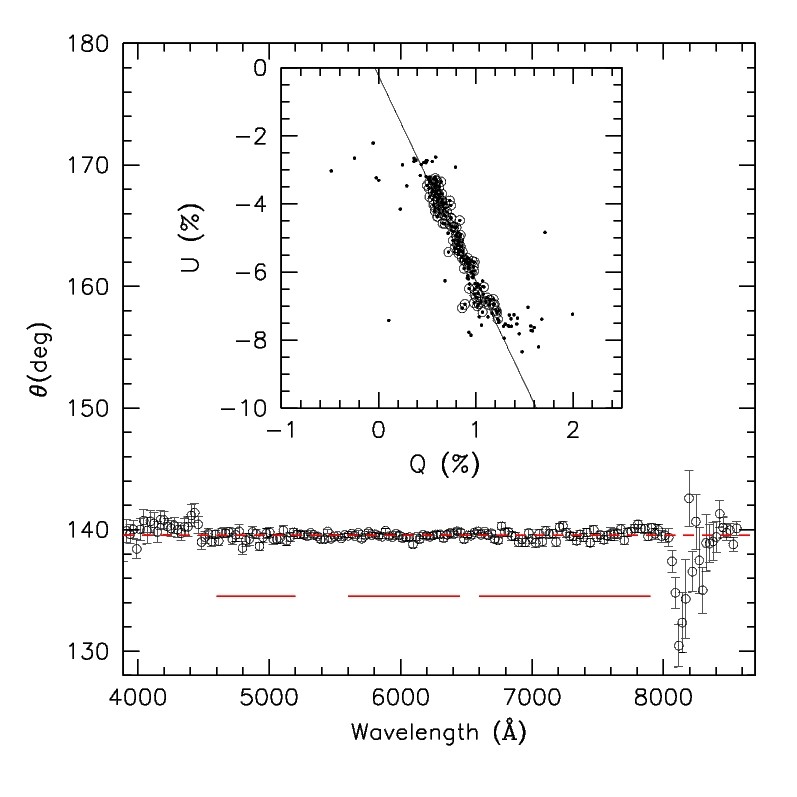}
\caption{\label{fig:ispang}Linear polarization position angle for the last
epoch. The dashed line is placed at the average value computed within
the wavelength ranges indicated by the three horizontal segments. The
insert shows the observed $Q$-$U$ plane. Circles mark data within the
selected wavelength ranges, while the straight line is a best fit to
this data subset.}
\end{figure}

Since the polarization position angle $\theta_{ISP}$ can be considered
wavelength independent (see for example Whittet et al. \cite{whittet};
see also Fig.~\ref{fig:isp}, lower panel), we have estimated it by
simply computing the average value within SN line-free regions in our
last-epoch spectropolarimetric data (see Fig.~\ref{fig:ispang}). This
turns out to be $\theta_{ISP}$=139.5$\pm$0.1 degrees. Using this
value, Eq.~\ref{eq:ispdeg} and Eq.~\ref{eq:isp}, the intrinsic SN
polarization can be derived by subtracting $Q_{ISP}$, $U_{ISP}$ from
the observed $Q$, $U$ Stokes parameters and recomputing the
polarization degree and polarization angle. This correction was
applied to all epochs, assuming that the ISP does not change with
time.  The constancy of the ISP is confirmed by inspection of the
observed spectropolarimetric data, which do not show any significant
variation in the line-free regions. This is illustrated more
quantitatively in Table~\ref{tab:isp}, where we have reported the
results of a linear squares fitting performed in the wavelength range
6400\AA$\leq \lambda
\leq$7200\AA, which is known to be free of strongly polarized SN lines
(Kasen et al. \cite{kasen04}; Leonard et al. \cite{leonard}; Wang et
al. \cite{wang06b}). Besides listing the slope $b$ and its RMS
uncertainty, Table~\ref{tab:isp} also presents the RMS scatter from the
fitted linear law ($\sigma$) and the polarization degree interpolated
at the center of the spectral range ($P_{6800}$) and its RMS
uncertainty. The small fluctuations seen in $P_{6800}$, which reach a
peak-to-peak value of 0.15\%, are fully consistent with a null
variation within the measured errors.

Since most of the extinction takes place in what is probably a cold
molecular cloud (Patat et al. \cite{patat07}; Cox \& Patat
\cite{cox}), we argue this is true for the ISP\footnote{The column 
densities associated with the time-variant features are two orders of
magnitude lower than that deduced for the saturated
components.}. During the epochs covered by our observations, that span
about 50 days, several narrow Na~I D features have shown a significant
evolution, which has been interpreted as due to changes in the
physical conditions of circumstellar material induced by the SN
radiation field and/or direct interaction with the fast expanding SN
ejecta (Patat et al. \cite{patat07}). In contrast, no variations have
been detected in the Na~I, Ca~II, CN, CH, CH$^+$ and Ca~I components
associated with the molecular cloud.

\begin{figure*}
\centering
\includegraphics[width=14cm]{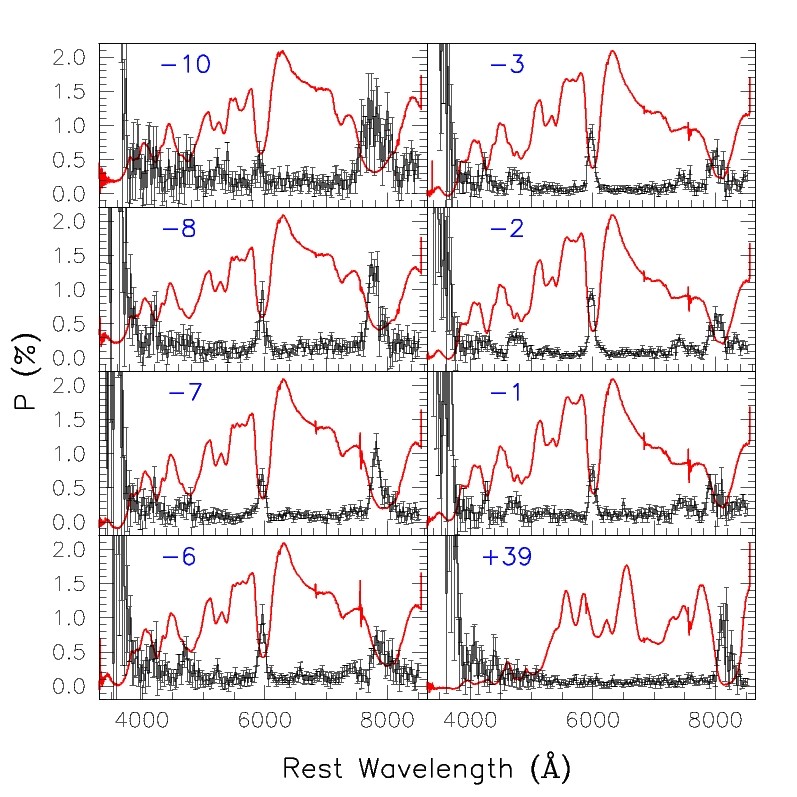}
\caption{\label{fig:polevol}Spectropolarimetric evolution of SN~2006X. For 
each epoch the ISP-corrected polarization degree is presented together
with the unbinned flux spectrum (solid line). Epochs refer to $B$ maximum
light.}
\end{figure*}

In principle, if the CSM material is rich in dust and provided that
dust can survive the SN radiation field, changes in its physical
conditions might be reflected in variations in the extinction suffered
by the SN. Of course, such changes are difficult to disentangle
from other effects on the photometry, because of the intrinsic SN
color evolution, the heavy reddening and the presence of a
light echo (Wang et al. \cite{wangx},
\cite{wangx08}; Crotts \& Yourdon \cite{crotts}). 
On the other hand, the ISP is so strong in SN~2006X in comparison to
what is expected for the intrinsic SN continuum polarization that the
effects of the SN evolution are negligible. This makes
spectropolarimetry a promising tool to reveal variations in the
conditions of the obscuring material.  As we have just shown, the ISP
degree appears to be constant within the measurement errors. This
suggests that the bulk of polarization takes place within the
molecular cloud, which is undisturbed by the SN explosion.

The derived ISP position angle appears to be tangential to the dust
lane associated to the spiral arm close to the explosion site (see
Fig.~\ref{fig:fchart}). This is perfectly in line with the well known
fact that dust grains tend to be aligned along the magnetic field,
which follows rather closely the spiral pattern of galaxies (see for
example Scarrot, Ward-Thompson \& Warren-Smith \cite{scarrot}). The
same behavior was seen, for instance, in the ISP deduced from the
spectropolarimetric data of SNIa 2001el (Wang et al. \cite{wang03}).

The fact that a light echo has been detected in SN~2006X (Wang et
al. \cite{wangx08}; Crotts \& Yourdon \cite{crotts}) suggests that an
additional, possibly time-dependent source of continuum polarization
might be present. In fact, if the distribution of the scattering dust
is asymmetric with respect to the line of sight, this is expected to
produce a net, non-null continuum polarization (Wang \& Wheeler
\cite{wang96b}; Patat \cite{patat05}) and broad polarized spectral features 
(Wang \& Wheeler \cite{wang96b}). However, while the polarized flux is
supposed to be high when the scattered light dominates over the light
directly emitted by the SN (i.e. at late epochs), that should not be
the case around maximum light, when the polarized photons are
extremely diluted into the basically unpolarized radiation emitted by
the photosphere. This additional effect can be quantified only by
detailed modeling.  Here we notice that since the ISP aligns with the
spiral pattern of NGC~4321, it is most probable that the bulk of
polarization arises in a cloud belonging to the disk of the host and
hence has nothing to do with the CS environment of SN~2006X.

\section{\label{sec:snpol}Intrinsic Polarization}

The spectropolarimetric evolution of SN~2006X on the eight epochs
covered by our observations is summarized in Fig.~\ref{fig:polevol},
(more detailed data for each single epoch are presented in
Figs.~\ref{fig:pol-10} to \ref{fig:pol+39}).

\begin{figure}
\centering
\includegraphics[width=8cm]{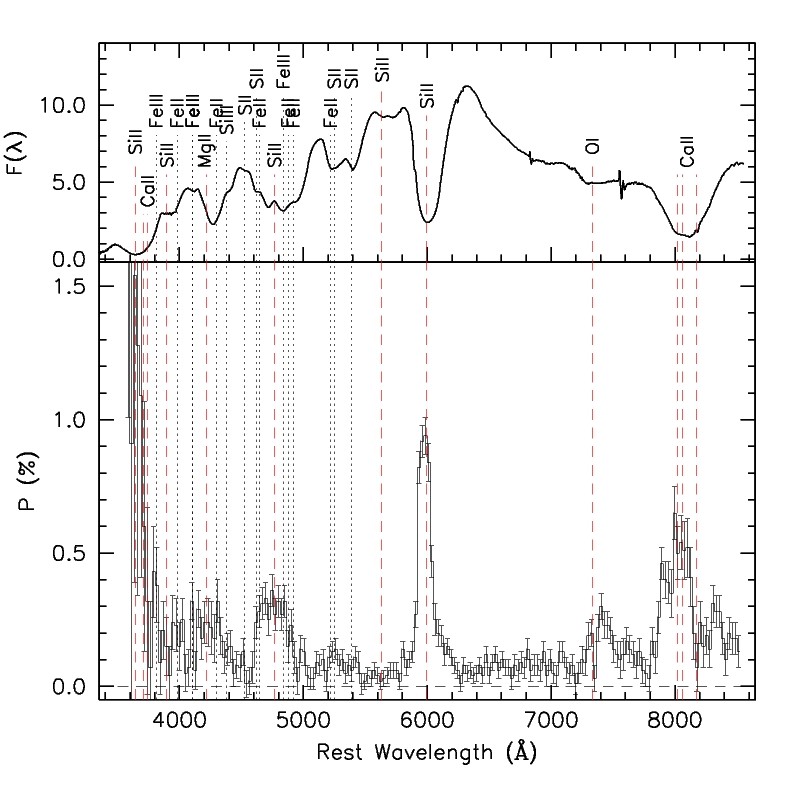}
\caption{\label{fig:polmax}Spectropolarimetry of SN~2006X on day $-$2.
Main line identifications are given in the upper panel. Vertical
dashed lines correspond to an expansion velocity of 17,000 km
s$^{-1}$, while the dotted lines correspond to 11,000 km s$^{-1}$.}
\end{figure}

As is generally the case for SNe (Cropper et al. \cite{cropper}), the
polarization signal is present in rough coincidence with the minima of
P-Cygni profiles. Moreover, as for other SNe~Ia (Wang et
al. \cite{wang03}), the most prominent features are observed in Si~II
6355\AA\/ and the near-IR Ca~II triplet. This is clearly
illustrated for day $-$2 in Fig.~\ref{fig:polmax}, where we have also
given tentative identifications for the main features, following the
case of SN~2004dt (Wang et al. \cite{wang06b}).

The evolution of most prominent features will be discussed in the next
subsections and here we just give a general overview of the
spectropolarimetric evolution.

Starting with the first spectrum, ten days before maximum light, the
most prominent feature is the Ca~II IR triplet, which shows a maximum
polarization as high as $\sim$1.5\%. A lower polarization signal is
detected for the Si~II 6344\AA\/ line, which displays a maximum
polarization of $\sim$0.5\%. Even though disturbed by a rather high
noise, other features in the blue spectral range, like the Si~II 5051\AA\/
and the Mg~II 4471\AA, exhibit low polarization levels
($\leq$0.5\%). Also, the continuum between 6200 and 7000\AA\/ shows a
low polarization level, of about 0.2\%, compatible with the
values reported for SNe 2001el (Wang et al. \cite{wang03}) and 2004dt
(Wang et al. \cite{wang06b}; Leonard et al. \cite{leonard}). As time
goes by, the Ca~II IR triplet polarization decreases and this is
accompanied by a quite rapid increase in the polarization level of the
Si~II 6355\AA\/ feature, which reaches a peak value of $\sim$1.1\% on
day $-$6 and decreases again to reach $\sim$0.8\% on day $-$1. This
roughly conforms to the behavior reported by Wang et
al. ({\cite{wang07}) for SN~2002bo, which reached the Si~II
polarization peak about at the same epoch.

\begin{table}
\caption{\label{tab:dominant}Dominant axis weighted fittings for the 
eight epochs.}
\tabcolsep 0.75mm
\begin{tabular}{ccccccc}
\hline
Phase  & $\alpha$ & $\beta$ & $\chi^2$ & $\sigma$ &$\theta_d$ & $p$\\
(days) & (\%)     &         &          & (\%)     & (deg) \\
\hline
$-$10  &$-$0.082(0.011) & $-$0.158(0.058) &   300 & 0.28 & 85.5(3.3)& 0.18\\               
  $-$8 &  +0.084(0.007) & $-$0.269(0.040) &   291 & 0.15 & 82.5(2.3)& $<$10$^{-8}$\\
  $-$7 &  +0.040(0.004) & $-$0.354(0.032) &   376 & 0.12 & 80.3(1.8)& $<$10$^{-8}$\\
  $-$6 &  +0.082(0.006) & $-$0.274(0.045) &   262 & 0.14 & 85.4(2.5)& $<$10$^{-8}$\\
  $-$3 &$-$0.002(0.004) & $-$0.404(0.035) &   514 & 0.14 & 79.0(2.0)& $<$10$^{-8}$\\
  $-$2 &  +0.042(0.003) & $-$0.362(0.027) &   706 & 0.13 & 80.1(1.5)& $<$10$^{-8}$\\
  $-$1 &$-$0.015(0.006) & $-$0.363(0.042) &   401 & 0.14 & 80.0(2.3)& $<$10$^{-8}$\\
+39  &  $-$0.003(0.004) & $-$0.018(0.063) &   275 & 0.16 & 89.5(3.6)& 0.73\\             
\hline
\multicolumn{7}{l}{Note: Data with 4000$\leq \lambda \leq$8600\AA\/ were used. The number of}\\
\multicolumn{7}{l}{degrees of freedom is 175.}
\end{tabular}
\end{table}

\begin{table}
\caption{\label{tab:domsi}Si~II dominant axis weighted fittings for the 
eight epochs.}
\tabcolsep 0.8mm
\begin{tabular}{ccccccc}
\hline
Phase  & $\alpha$ & $\beta$ & $\chi^2$ & $\sigma$ &$\theta_d$  & $p$\\
(days) & (\%)     &         &          & (\%)     & (deg)      &    \\
\hline
$-$10  &  +0.024(0.042) & $-$0.192(0.213) &   19 & 0.20 & 84.6(12.1)& 0.55\\
  $-$8 &  +0.131(0.034) & $-$0.625(0.162) &   22 & 0.15 & 74.0(8.6)& 8$\times$10$^{-3}$\\
  $-$7 &  +0.099(0.019) & $-$0.509(0.095) &   21 & 0.11 & 76.5(5.2)& 9$\times$10$^{-3}$\\
  $-$6 &  +0.110(0.032) & $-$0.687(0.124) &   13 & 0.12 & 72.8(6.5)& 2$\times$10$^{-4}$\\
  $-$3 &  +0.043(0.021) & $-$0.576(0.074) &   81 & 0.20 & 75.0(4.0)& 0.02\\
  $-$2 &  +0.115(0.015) & $-$0.401(0.050) &  141 & 0.21 & 79.1(2.8)& 0.04\\
  $-$1 &  +0.064(0.021) & $-$0.443(0.095) &   54 & 0.18 & 78.1(5.2)& 0.03\\
+39  &  $-$0.010(0.012) & $-$0.318(0.235) &   14 & 0.05 & 81.2(13.1)& 0.47\\               
\hline
\multicolumn{7}{l}{Note: Data with velocities between $-$25,000 and $-$10,000 km s$^{-1}$}\\
\multicolumn{7}{l}{were included. The number of degrees of freedom is 11.}
\end{tabular}
\end{table}

On day $-$1 the Ca~II polarization appears to be reduced to
0.63$\pm$0.17. Finally, on the last epoch (day +39), no trace of Si~II
polarization is seen and the only detectable lines are the Ca~II IR
triplet ($\sim$1\%) and some unidentified features between 4000 and
6000\AA. The continuum in the 6000-7000\AA\/ region shows a very low
polarization level (0.1\%) which, given the measurement errors, is
consistent with a null polarization and lends some support to the
applied ISP correction (see Sect.~\ref{sec:isp}). The apparent {\it
re-polarization} of the Ca~II IR triplet detected on day +39 will be
discussed in Sect.~\ref{sec:ca}.

In general, SN~2006X has an intermediate behavior between SN~2001el
(Wang et al. \cite{wang03}), for which the Ca~II IR triplet was
more polarized than the Si~II 6355\AA\/ line, and SN~2004dt (Wang et al.
\cite{wang06b}), which showed the opposite. Also, the latter object
has shown a lower continuum polarization level, while the former was
more polarized than SN~2006X. Despite the strong signal detected in
several spectral features, SN~2004dt did not show any trace of
polarization corresponding to the O~I 7774 line (Wang et al.
\cite{wang06b}; Leonard et al. \cite{leonard}). This is also the case for 
SN~2006X (see Figs.~\ref{fig:polevol}, \ref{fig:polmax} and
Sect.~\ref{sec:o}), for which the O~I polarization is certainly lower
than 0.3\% at any epoch.

\begin{figure*}
\centering
\includegraphics[width=14cm]{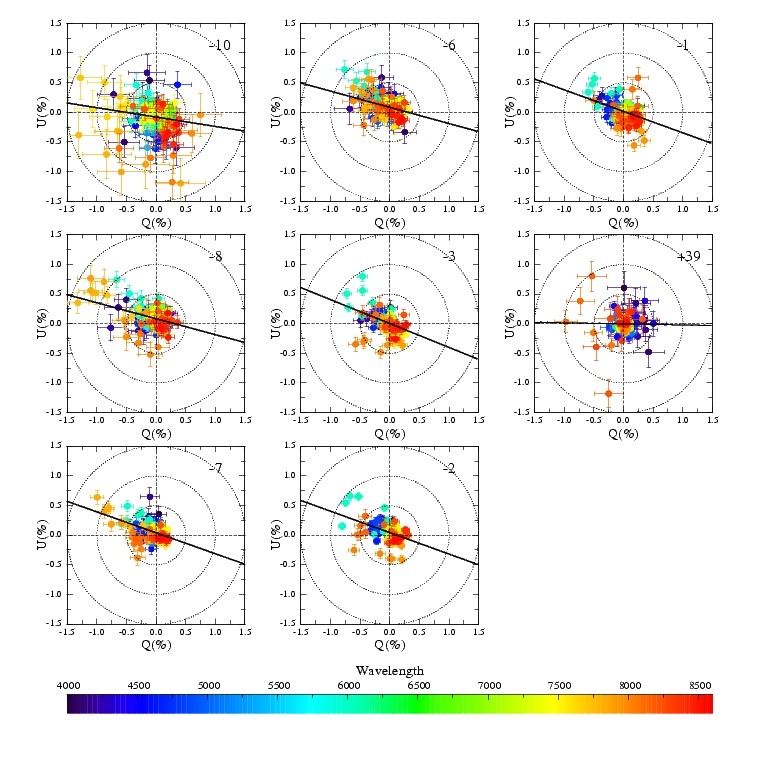}
\caption{\label{fig:dominant}ISP-corrected spectropolarimetric data of SN~2006X
on the $Q$-$U$ plane as a function of wavelength for all epochs
(indicated on the upper right corner of each panel). The straight line
traces the dominant axis derived from a weighted least squares fit
(see Sect.~\ref{sec:snpol}), while the three dotted circles indicate
0.5, 1.0 and 1.5\% polarization levels.}
\end{figure*}

Before proceeding with the detailed study of the most prominent
features, it is interesting to analyze the polarization on the
plane of the Stokes $Q$-$U$ parameters. The presence of a
dominant direction can often be identified through the geometric
distribution of points in this plane (see for example Wang et
al. \cite{wang01}; Wang \& Wheeler \cite{WW08}).  
For this purpose, following Wang et
al. (\cite{wang03}), we have performed a weighted linear squares fit
to the data using the linear relation $U=\alpha\; +
\beta\; Q$.  The position angle of the dominant
axis is then given by $\theta_d=1/2 \; \arctan \beta$.  We have done
this for each epoch in the wavelength range 4000$\leq \lambda
\leq$8600\AA.  The results are plotted in Fig.~\ref{fig:dominant} and
summarized in Table~\ref{tab:dominant}, where we have also reported
the uncertainties on $\alpha$, $\beta$ and $\theta_d$, the RMS
deviation $\sigma$ from the best fit relation and the probability $p$
of chance linear correlation (low values of $p$ indicate a
significant correlation). Of course, the relation between $\theta_d$
and $\beta$ rigorously holds only if the best fit axis goes through
the origin of the $Q$-$U$ coordinates system.  As one can see from the
values of $\alpha$ in Table~\ref{tab:dominant}, this is true to a good
approximation (see also Fig.~\ref{fig:dominant}).

The first thing to notice is that on the first and on the last epoch
there is no clear dominant axis (the probability of chance correlation
is high, see Tab.~\ref{tab:dominant}). On day $-$10 this is probably
due to the relatively large errorbars on the data, while on day +39
this is most likely related to the intrinsic low polarization.  With
the exception of these two cases, a dominant axis seems to be present,
almost parallel to the $Q$ axis: the weighted average position angle
is $<\theta_d>$=+82.0$\pm$0.8 degrees. Even though the data points
appear to get closer and closer to the dominant axis as the SN
approaches maximum light, significant departures are visible.
SN~2006X would be classified as spectropolarimetry type D1 in the
terminology of Wang \& Wheeler (\cite{WW08}) corresponding to a
distinct dominant axis, but also significant departures from that
axis. Individual lines would be classed as SP Type L for loops, as
discussed in the next sections.

\subsection{\label{sec:si}The Si~II 6355\AA\/ line}

At the first epoch, the polarization of Si~II 6355\AA\/ line peaks at
$\sim$0.57$\pm$0.19\% (see Fig.~\ref{fig:sievol}).  In the subsequent
days the polarization degree increases and the polarization peak gets
narrower and, at variance with the Ca~II IR triplet (see next
section), it is only very slightly blue-shifted with respect to the
line absorption minimum. Maximum peak polarization is reached on day
$-$6 (1.05$\pm$0.14\%), and in the following days it slightly decreases,
to reach 0.75$\pm$0.10\% on day $-1$. Finally, on our last epoch (+39)
the polarization is $\leq$0.15\%.

A dominant axis is clearly present for the Si~II line on day $-$7 (see
Fig.~\ref{fig:siprof}, upper panel), at an average position angle
$\theta$=70.4$\pm$8.2 degrees. This value, which was computed in the
velocity range $-$23,000 $\leq$v$\leq$ $-$17,000 km s$^{-1}$, is
consistent at the 1-$\sigma$ level with the one deduced for Ca~II
(78.0$\pm$3.4; see next section) and at the 2-$\sigma$ level with the
overall angle as given in Table~\ref{tab:dominant} and illustrated in
Fig.~\ref{fig:dominant}. Following Wang et al. (\cite{wang03}), we
computed the components of the Si~II polarization along ($P_d$) and
perpendicular ($P_o$) to the dominant axis, which are obtained by
rotating counterclockwise the observed $Q$-$U$ coordinates system by
the angle $\theta$ that defines the dominant axis.

The Si~II polarization is practically parallel to the dominant axis,
i.e. $P_o\approx$0 (see Fig.~\ref{fig:siprof}, lower panel), at
variance with what is seen for Ca~II (see next section). A hint of a
dominant axis is still present on day $-1$, at least for the core of
the absorption profile and with a lower value for the position angle
(see Fig.~\ref{fig:siprof2}). Nevertheless, $Q$-$U$ diagrams show
significant deviations and the presence of clear loops cut through by
the dominant axis (Fig.~\ref{fig:dom_si}).

The results of a weighted linear squares fit in the Si~II $Q$-$U$ plane
are presented in Table~\ref{tab:domsi} (see also
Sect.~\ref{sec:snpol}).

\begin{figure}
\centering
\includegraphics[width=8cm]{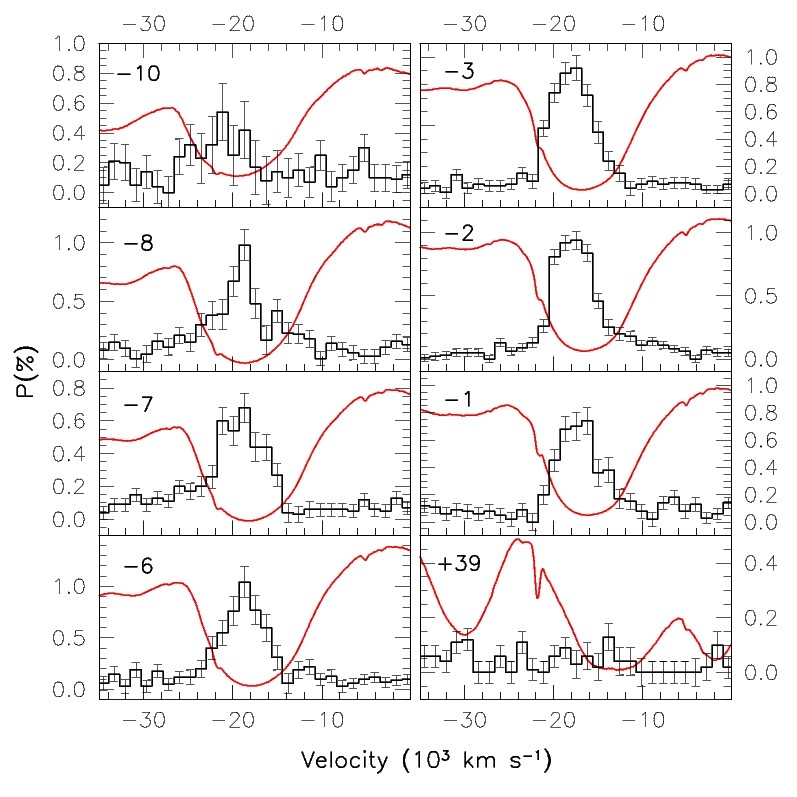}
\caption{\label{fig:sievol}Polarimetric evolution of the Si~II 6355\AA\/
line profile. The smooth solid curve traces the unbinned flux
spectrum, arbitrarily scaled for presentation. Note the different
vertical scales in each plot. The weak feature around $-$22,000 km
s$^{-1}$ is due to IS Na~I D absorption.}
\end{figure}

\begin{figure}
\centering
\includegraphics[width=8cm]{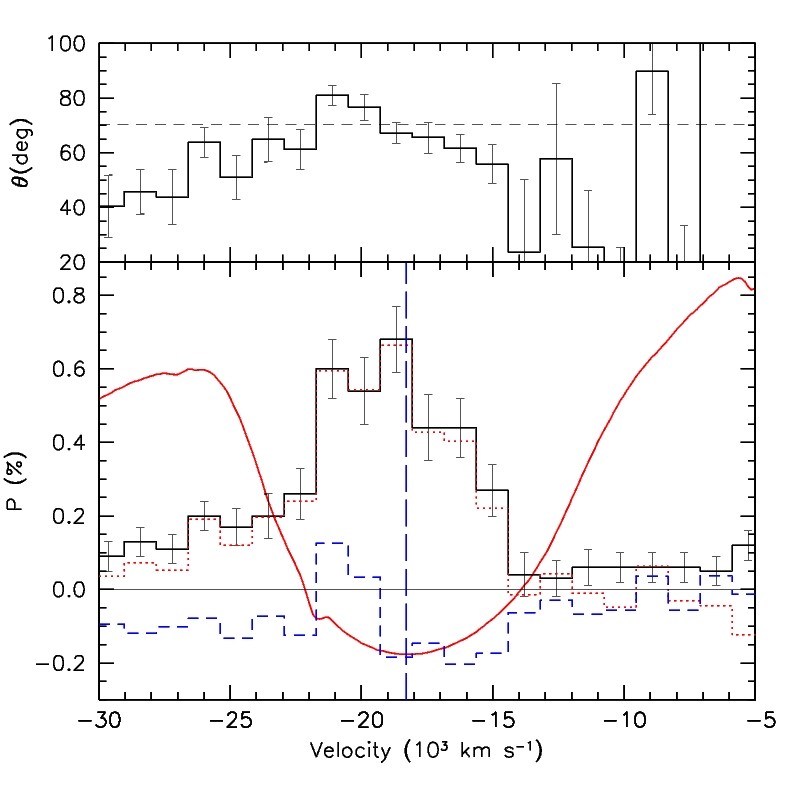}
\caption{\label{fig:siprof}Lower panel: polarization profile of Si~II 6355\AA\/
on day $-$7. Dotted and dashed lines trace $P_d$ and $P_o$
respectively (see Sect.~\ref{sec:si}). The vertical dashed line
indicates the expansion velocity deduced from Si~II 6355\AA\/
absorption minimum for this epoch. Upper panel: polarization position
angle. The horizontal dashed line indicates the average angle computed
in the velocity range $-$23,000 $\leq$v$\leq$ $-$17,000 km
s$^{-1}$.The weak absorption feature around $-$22,000 km s$^{-1}$ is
due to IS Na~I D.}
\end{figure}

\subsection{\label{sec:ca}The Ca~II IR triplet}

As in the case of the well studied SN~2001el (Wang et
al. \cite{wang03}), the Ca~II IR triplet is the most prominent feature
seen during the very early phases. Its polarimetric evolution is
presented in Fig.~\ref{fig:caevol}. As already mentioned in
Sect.~\ref{sec:evol}, this feature shows extremely high expansion
velocities, giving evidence of material moving as fast as 40,000 km
s$^{-1}$ on day $-$10 (see Fig.~\ref{fig:ca}). At this epoch, the
minimum of the absorption trough is at about 27,000 km s$^{-1}$,
which is much higher than the expansion velocity deduced for the
Si~II 6355\AA\/ absorption minimum at the same epoch, i.e. 19,500 km s$^{-1}$.

Most likely the Ca~II line profile is a signature of a HVF (Mazzali et
al. \cite{mazzali05a}, Gerardy et al. \cite{gerardy}, and references
therein), of the type discovered in SN~2001el (Wang et
al. \cite{wang03}), even though in that case the HVF was clearly
detached from the photospheric component (Wang et al. \cite{wang03};
Kasen et al. \cite{kasen}).  Due to the much higher expansion
velocities measured in SN~2006X, the line broadening causes a
significant blending. In general, the polarization profile is
non-homomorph with respect to the flux profile, with the polarization
peak being systematically bluer than the absorption center. This is
better seen in Fig.~\ref{fig:caprof}, where we present the data for
day $-$7. The polarization peaks at about $-$26,000 km s$^{-1}$
(1.18$\pm$0.11\%) while the expansion velocity deduced from Si~II
6355\AA\/ on the same epoch is $\sim$16,600 km s$^{-1}$. The
photospheric component of Ca~II shows a polarization signal, that is
definitely lower than in the HVF (0.34$\pm$0.11\%). As shown by the
upper panel of Fig.~\ref{fig:caprof}, the position angle within the
HVF shows relatively small variations: the average value computed in
the velocity range $-$30,000 $\leq$v$\leq$ $-$24,000 km s$^{-1}$ is
$<\theta>$=+78.0$\pm$3.4 degrees, which is actually not very far from
the dominant axis position deduced from the whole data sets
(82.0$\pm$0.8). The photospheric component shows no clear dominant
axis (see Fig.~\ref{fig:caprof}).

The result of the $P_d$-$P_o$ decomposition is shown in the lower
panel of Fig.~\ref{fig:caprof}. As expected, $P_o\approx$0 within the
HVF, while $P_d$ and $P_o$ contribute in a similar way to the
polarization of the photospheric component, suggesting that the matter
concerned is clumpy (Wang et al. \cite{wang03}; Tanaka et al. \cite{tanaka}).

Another interesting fact is the rapid time evolution. As the SN
approaches maximum light things change rather dramatically. On day
$-$2 (see Fig.~\ref{fig:caprof2}) the minimum of the absorption trough
corresponds to a velocity which is very close to the one indicated by
the Si~II 6355\AA\/ feature ($-$16,600 km s$^{-1}$). The polarization
still peaks at higher velocities, even though the peak is not very
well defined and two different features seem to co-exist.  One,
centered at about $-$18,000 km s$^{-1}$ (0.7$\pm$0.1\%), shows a
dominant axis, not too far from the one identified in the previous
epoch and actually very close to the average dominant axis defined by
the whole data set. The higher velocity component, peaking around
$-$24,000 km s$^{-1}$, has a slightly lower polarization
($\sim$0.5$\pm$0.1\%) but a definitely different orientation: even though
the position angle changes gradually across the feature
(Fig.~\ref{fig:caprof2}, upper panel), on average it is practically
orthogonal to the slower component.  This is shown also by the
$P_d$-$P_o$ decomposition (Fig.~\ref{fig:caprof2}, lower panel).

In general, the Ca~II polarization shows significant deviations from a
dominant axis. This is illustrated in Fig.~\ref{fig:dom_ca}, where we
present $Q$-$U$ diagrams for all epochs, including data in the
velocity range $-$40,000 $\leq$v$\leq$ $-$10,000 km s$^{-1}$. With
only the exception of the first epoch, the best-fit axis is never too
far from the average dominant axis, even though pronounced loops and a
quite erratic behavior are clearly seen. The results of a weighted
linear squares fit in the $Q$-$U$ plane are presented in
Table~\ref{tab:domca} (see also the introductory part of
Sect.~\ref{sec:snpol}).

\begin{table}
\caption{\label{tab:domca}Ca~II dominant axis weighted fittings for the 
eight epochs.}
\tabcolsep 0.6mm
\begin{tabular}{ccccccc}
\hline
Phase  & $\alpha$ & $\beta$ & $\chi^2$ & $\sigma$ &$\theta_d$  & $p$\\
(days) & (\%)     &         &          & (\%)     & (deg)      & \\
\hline
$-$10  &$-$0.224(0.045) & $-$0.264(0.095) &   59 & 0.43 & 82.6(5.4)& 0.06\\
  $-$8 &  +0.041(0.025) & $-$0.335(0.058) &   55 & 0.24 & 80.7(3.3)& 3$\times$10$^{-5}$\\
  $-$7 &$-$0.012(0.015) & $-$0.345(0.048) &   89 & 0.17 & 80.5(2.7)& 3$\times$10$^{-5}$\\
  $-$6 &  +0.038(0.021) & $-$0.293(0.080) &   59 & 0.19 & 81.9(4.5)& 0.02\\
  $-$3 &$-$0.103(0.013) & $-$0.157(0.088) &  129 & 0.19 & 85.6(5.0)& 0.82\\
  $-$2 &$-$0.078(0.011) & $-$0.190(0.067) &  143 & 0.17 & 84.6(3.8)& 0.11\\
  $-$1 &$-$0.105(0.021) & $-$0.298(0.119) &   82 & 0.22 & 81.7(6.7)& 0.33\\
+39  &  $-$0.030(0.010) &   +0.032(0.106) &   88 & 0.30 & 90.9(6.1)& 0.62\\
\hline
\multicolumn{7}{l}{Note: Data with velocities between $-$40,000 and $-$10,000 km s$^{-1}$}\\
\multicolumn{7}{l}{were included. The number of degrees of freedom is 31.}
\end{tabular}
\end{table}

Finally, there is an interesting fact that concerns the last epoch of
our data set. At variance with the well studied SN~2001el (Wang et
al. \cite{wang03}), which had shown no detectable trace of
polarization in the Ca~II IR triplet on day +41 (bearing in mind that
the signal-to-noise ratio was lower in that case), and SN~2004du
that displayed a very low polarization level ($\sim$0.2\%) 18 days
after maximum light (Leonard et al. \cite{leonard}), SN~2006X presents
a significant signal at a comparably late epoch (day +39). The
polarized feature, centered at about $-$16,000 km s$^{-1}$, reaches
1.20$\pm$0.24\% (see Fig.~\ref{fig:caprof3}). Even though the SN was
already at $V\sim$16 at these phases (see Table~\ref{tab:obs}), as a
result of the 2 hours total integration time the signal-to-noise ratio
in the combined binned flux spectrum is about 340 at the minimum of
the Ca~II absorption trough (see Fig.~\ref{fig:isp}, upper plot), so
that the detected polarized feature, reaching a 5-$\sigma$ level, is
real. The data do not show evidence for a dominant axis; on the
contrary, the polarization angle shows a smooth variation across the
absorption (Fig.~\ref{fig:caprof3}), which turns into a loop
on the $Q$-$U$ plane (Fig.~\ref{fig:dom_ca}). An important aspect to
note is that the high-velocity component visible on day $-$1 at about
$-$24,000 km s$^{-1}$ (Fig.~\ref{fig:caevol}) has completely
disappeared on day +39, while the other ($-$16,000 km s$^{-1}$) has
grown in polarization and possibly receded in velocity.  As one can
see in Fig.~\ref{fig:caprof3}, the emergence of this polarized line is
accompanied by a substantial broadening of the absorption trough in
the flux spectrum on the red side.  A lower polarization ($<$0.5\%) is
seen also around $-$10,000 km s$^{-1}$ but, given the uncertainties,
this is only marginally significant and is consistent with a null
polarization at the $\sim$2.5-$\sigma$ level.

\begin{figure}
\centering
\includegraphics[width=8cm]{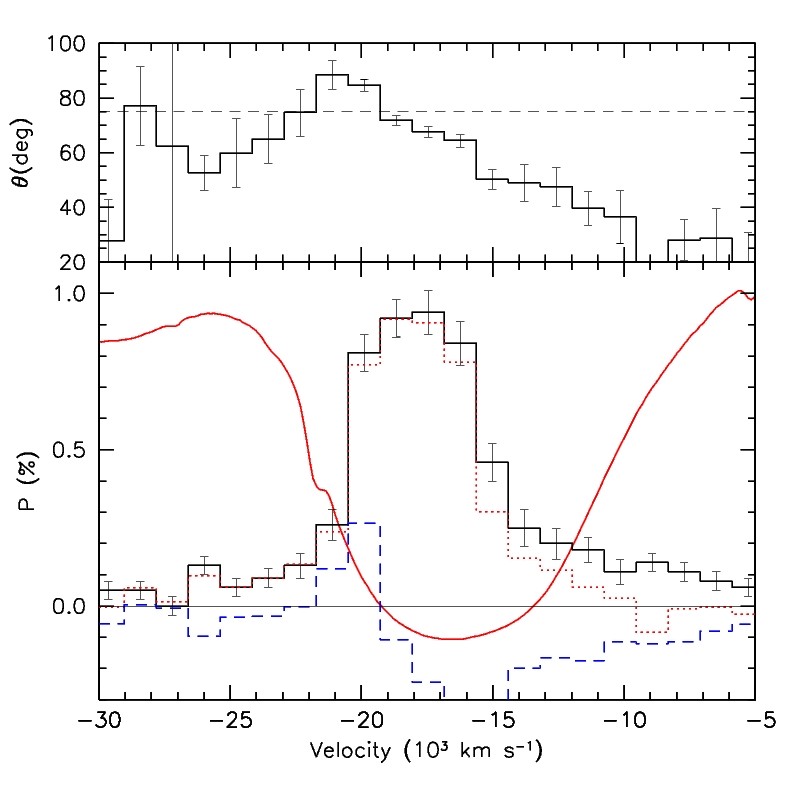}
\caption{\label{fig:siprof2}Same as Fig.~\ref{fig:siprof} for day $-$2.
The horizontal dashed line in the upper panel indicates the dominant
axis deduced on day $-$7.}
\end{figure}

\subsection{\label{sec:o}The O~I 7774\AA\/ line and other minor features}

\begin{figure*}
\centering
\includegraphics[width=14cm]{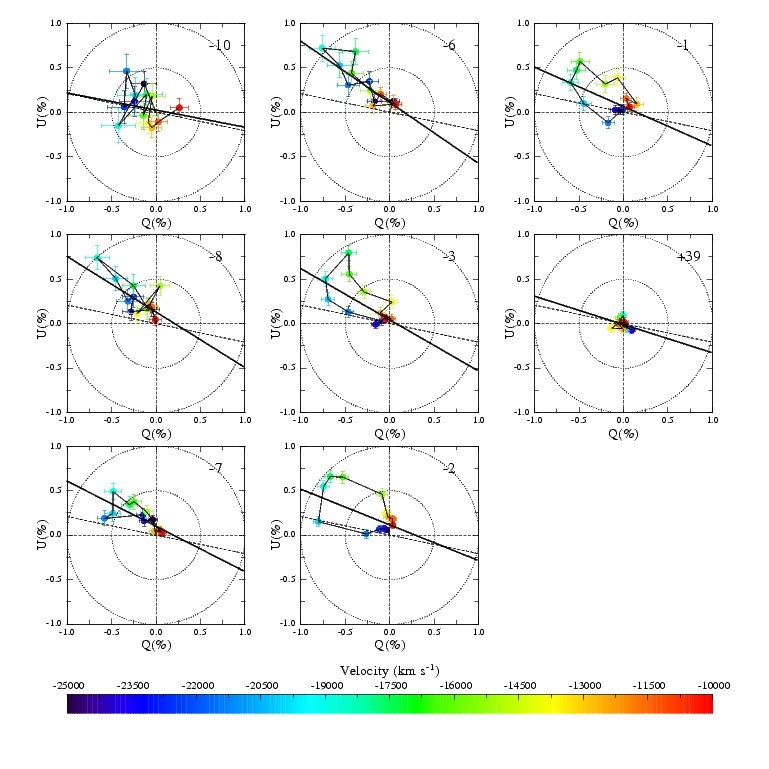}
\caption{\label{fig:dom_si}$Q$-$U$ diagrams for Si~II 6355\AA. The dotted
line traces the average dominant axis at $<\theta_d>$=+82.0 degrees
(see Sect.~\ref{sec:snpol}), while the solid line is the linear
weighted fit to the displayed Si~II data.}
\end{figure*}

The O~I 7774\AA\/ does not show any detectable polarization (see
Fig.~\ref{fig:oevol}), as in SN~2004dt (Wang et al. \cite{wang06b}).
At all epochs the polarization across this line is $\leq$0.2\%.  Wang
et al. (\cite{wang05}, \cite{wang06b}) have proposed that primordial
oxygen is more ``spherically'' distributed than newly-synthesized
silicon, magnesium and calcium. An important part of the argument for
the case of SN~2004dt was that the O~I and Si~II had closely the same
velocity profile, i.e. they shared the same velocity space and yet had
substantially different polarization. This conclusion holds also for
SN~2006X, at least in the earliest epochs, when the O~I line can be
clearly traced (Fig.~\ref{fig:osi}).

In the framework of hydrodynamical and nucleosynthesis models, the
difference in polarization between O and Si/Ca may be understood as a
consequence of their sources (see Fig. \ref{fig:DDT}). Oxygen
originates from both pre-explosion unburned WD matter in the outer
layers and from freshly synthesized O during explosive C-burning. The
corresponding O abundances are similar, namely $\approx$50\% and 70\%,
respectively. Whereas the former, dominant component can be expected
to be close to spherical, the latter will show the imprint of the
explosion.  In contrast, Si/Ca are mostly produced during the
explosion and reflect the explosion properties. As discussed by
Hoeflich et al. (\cite{hoeflich06}) within the framework of
off-centered delayed detonation models developed for SN 2004dt, O
shows no (or very weak) polarization because the pre-explosion O is
already sufficient to block all direct light emitted from the
photo-disk (the photosphere cross section along the line of
sight). This means there is no large scale, chemical asymmetry at the
wavelengths corresponding to the blue-shifted O absorption.

Besides the Si~II 6355\AA\/ and the near-IR Ca~II triplet, the only
spectral features for which polarization is detected at a significant
level are Si~II 5051\AA\/ and, less markedly, Mg~II 4471\AA\/ (see
Figs.~\ref{fig:pol-10}-\ref{fig:pol-1}). Fig.~\ref{fig:si2evol}
illustrates the evolution of Si~II 5051\AA. While on days $-$10 and
$-$8 the signal-to-noise is too low, on day $-$6 the polarization of
this line is detected at the 3-$\sigma$ level (0.37$\pm$0.11\%).  On
day $-$1 the feature is still there (0.36$\pm$0.06\%), while by day
+39 it has completely disappeared below the detection limit
($\leq$0.2\%). In general, the polarization peak appears to be broader
than that of Si~II 6355\AA\/ (see Fig.~\ref{fig:sievol}) and this
might be due to the presence of some other lines, whose absorption
troughs are clearly visible in the flux spectrum
(Fig.~\ref{fig:si2evol}). Less clear is the case of Mg~II 4471\AA.
Even though a polarized feature is consistently present from day $-$8
on, the detection level is significant only in the epochs immediately
preceding maximum light, when the polarization reaches 0.5$\pm$0.15\%
(see Figs.~\ref{fig:polm3}, \ref{fig:polm1}). In general, the $Q$-$U$
diagrams for these two weak features are quite noisy and it is very
difficult to say whether a dominant axis exists or not.

\subsection{\label{sec:cont}The 6400-7200\AA\/ continuum}

The region between 6400 and 7200\AA\/ is practically free of strong
features (Kasen et al. \cite{kasen04}; Leonard et al. \cite{leonard};
Wang et al. \cite{wang06b}) and can therefore be used to estimate the
continuum polarization. In the case of SN~2006X, the average
polarization level computed in this spectral range is below 0.2\% at
all epochs, which is consistent with a very small departure from
spherical symmetry in the photosphere ($\leq$10\%; H\"oflich
\cite{hoeflich}), even though fluctuations at the level of 3-4$\sigma$
from the average value are seen (see
Fig.~\ref{fig:contevol}). Finally, the analysis of the $Q$-$U$ plane for
the data in the wavelength range 6400\AA$\leq
\lambda \leq$ 7000\AA\/ does not show any significant dominant axis.
Incidentally, this confirms that the ISP was properly removed by the
procedure outlined in Sect.~\ref{sec:isp}.

\section{\label{sec:disc}Discussion and conclusions}

SN~2006X is characterized by a number of distinguishing features: a)
strong reddening and peculiar extinction law; b) extremely high
expansion velocities; c) intense high-velocity Ca~II features; d)
possible detection of circumstellar material; e) detection of a light
echo. These facts have led Wang et al. (\cite{wangx}) to propose that
SN~2006X and other rapidly expanding objects (like SN~2002bo) belong
to a subclass with distinct photometric and spectroscopic properties,
possibly associated with dusty environments. A similar proposal was
put forward for SN~2004dt by Wang et al. (\cite{wang06b}), who
suggested that these events might come from ``younger'' progenitor
stars.

\begin{figure}\centering
\includegraphics[width=8cm]{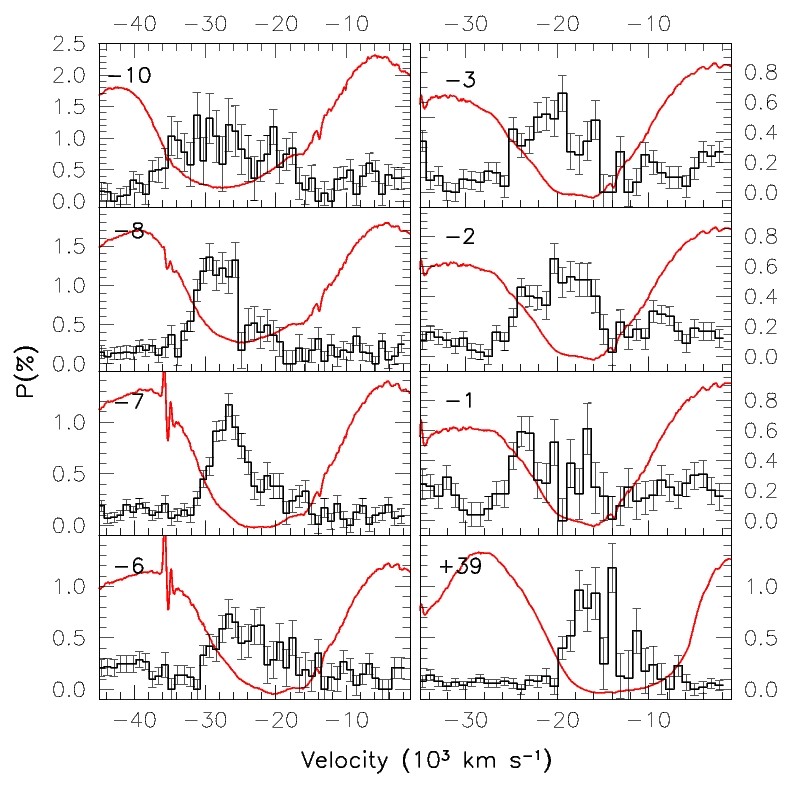}
\caption{\label{fig:caevol}Polarimetric evolution of the Ca~II IR 
triplet profile.  The smooth solid curve traces the unbinned flux
spectrum, arbitrarily scaled for presentation. Note the different
vertical and horizontal scales in each plot.}
\end{figure}

Also from a spectro-polarimetric point of view SN~2006X shows a number
of distinguishing features, each of them deserving a separate
discussion.

\begin{figure}
\centering
\includegraphics[width=8cm]{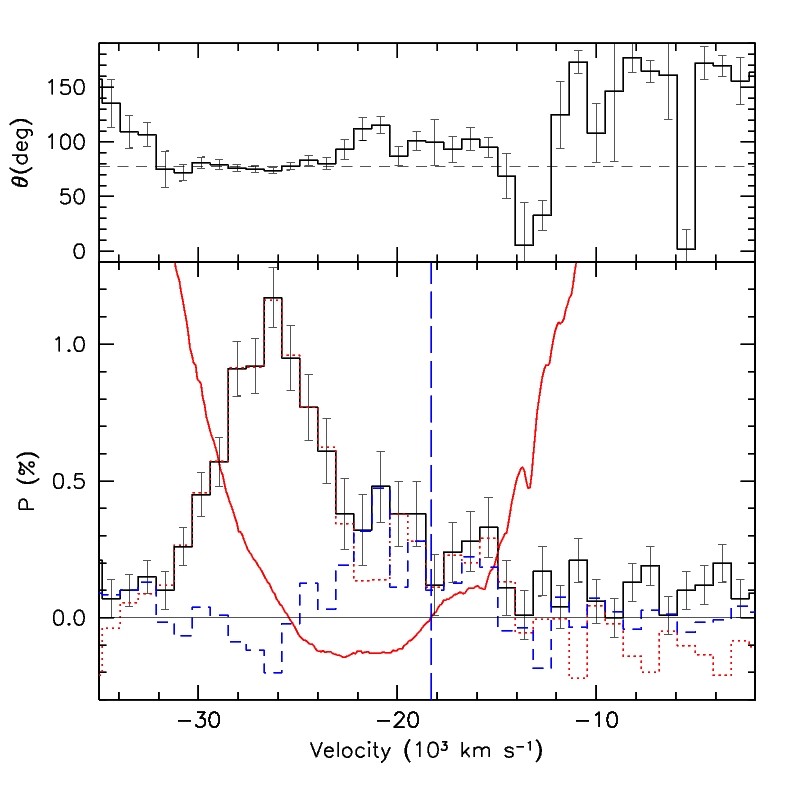}
\caption{\label{fig:caprof}Lower panel: polarization profile of 
Ca~II IR triplet on day $-$7. Dotted and dashed lines trace $P_d$ and
$P_o$ respectively (see Sect.~\ref{sec:si}). The vertical long-dashed
line indicates the expansion velocity deduced from the Si~II
6355\AA\/ line at the same epoch.  Upper panel: polarization position
angle. The horizontal dashed line indicates the average angle computed
in the velocity range $-$30,000 $\leq$v$\leq$ $-$24,000 km
s$^{-1}$. The feature around $-$15,000 km s$^{-1}$ is due to a
telluric absorption.}
\end{figure}

\subsection{\label{sec:gen}General spectropolarimetric properties}

Besides SN~2001el, SN~2006X is the only Type Ia SN with a very good
spectropolarimetric coverage during the pre-maximum phase and with
data extending up to more than one month past maximum. This has
allowed us to study the variations shown by the most prominent
features. In particular, the polarization of the Si~II 6355\AA\/ line
appears to follow the trend displayed by the fast expanding SN~2002bo
(Wang et al. \cite{wang07}), as shown in Fig.~\ref{fig:sipol}.  In a
scenario where the polarization across the absorption profiles is
generated by clumps each of them having a significant photospheric
covering factor (Wang et al. \cite{wang03}; Kasen et
al. \cite{kasen}), the observed fluctuations might be explained in
terms of changes in the covering factor itself as the photosphere
recedes in velocity with time.  Nevertheless, given the relatively
large errorbars, it is not possible to deduce a systematic
physical difference between the two events (e.g. different extension
and number of clumps in the ejecta of the two objects). Only future
observations with higher accuracy will enable this kind of analysis
for newly discovered, nearby SNe.

\begin{figure}
\centering
\includegraphics[width=8cm]{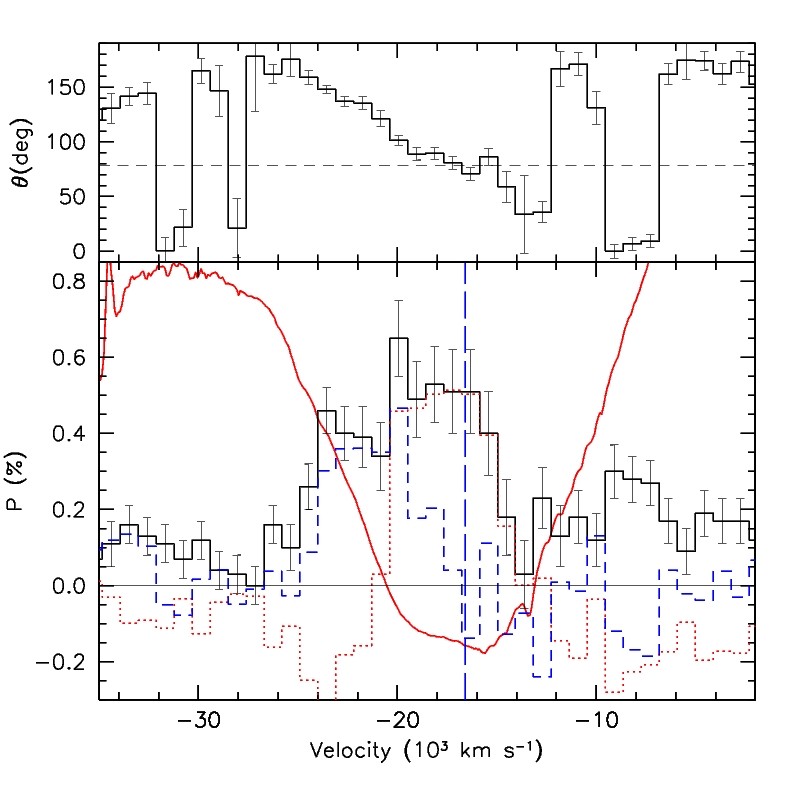}
\caption{\label{fig:caprof2}Same as Fig.~\ref{fig:caprof} for day $-$2.
The horizontal dashed line in the upper panel indicates the
orientation of the dominant axis deduced on day $-$7.}
\end{figure}

As shown by Wang et al. (\cite{wang07}), there is an anti-correlation
between the luminosity of Type Ia SNe and their degree of
asphericity. In order to properly place our object into the Wang et
al. (\cite{wang07}) plot, we have estimated the Si~II 6355\AA\/
polarization degree $P_{SiII}$ on day $-$5 using a second order
polynomial fitting to the observed data (1.0\%; see
Fig.~\ref{fig:sipol}, dashed curve). As for the error-bar, we have
adopted the same value estimated for the observed points at comparable
epochs (0.1\%). Once this is done, SN~2006X conforms to the behavior
shown by the other SN Ia, as we show in Fig.~\ref{fig:poldm} (after
excluding the two most deviating SNe 1999by and 2004dt the Pearson
correlation coefficient is 0.85, and the rms deviation of the data
points from the best fit is 0.14\%). Interestingly, such a relation
had been predicted within the delayed detonation models, where initial
chemical inhomogeneities, in the form of Ni clumps in the Si/O rich
outer layers, are expected to be produced during the deflagration
phase (Khokhlov \cite{khokhlov95}; Lisewski et al. \cite{lisewski00};
Gamezo et al. \cite{gamezo03}). However, the subsequent detonation
will burn most of the intermediate mass elements to $^{56}$Ni and,
thus, eliminate most of the chemical inhomogeneities at an efficiency
which increases with the total production of $^{56}$Ni, i.e. with the
luminosity (Khokhlov \cite{khokhlov01}; H\"oflich et
al. \cite{hoeflich02}; Wang et al. \cite{wang07}; Mazzali et
al. \cite{mazzali07}; Kasen et al. \cite{kasen09}).

SN~2006X shows one of the highest polarizations ever observed in a Ia
and it lies at the upper edge of the Wang et al. (\cite{wang07})
relation. As in all other known cases, the Si~II polarization
disappears after maximum light (see Fig.~\ref{fig:sievol}).

The line polarizations discussed here can be understood in a simple
topological picture, which has been confirmed by detailed models.
Thomson scattering produces null polarization in the case of forward
scattering, while the polarization is maximum if the photons are
scattered by 90 degrees. Consequently, the resulting polarization of
an individual element of the photo-disk increases with its distance
from the center.  Polarization by lines can be understood as follows:
a line "blocks out" at certain wavelengths the (polarized) flux from
the Thomson scattering dominated photo-disk. If the amount of blocking
is not distributed evenly around the line of sight, the result is that
the integrated polarization over the entire disk is not null (Wang et
al. \cite{wang03}; Kasen et al. \cite{kasen}; Wang et
al. \cite{wang06b}; H\"oflich et al. \cite{hoeflich06}). Uneven
blocking/optical depths can be caused by asymmetric chemical and/or
density distributions.

\begin{figure}
\centering
\includegraphics[width=8cm]{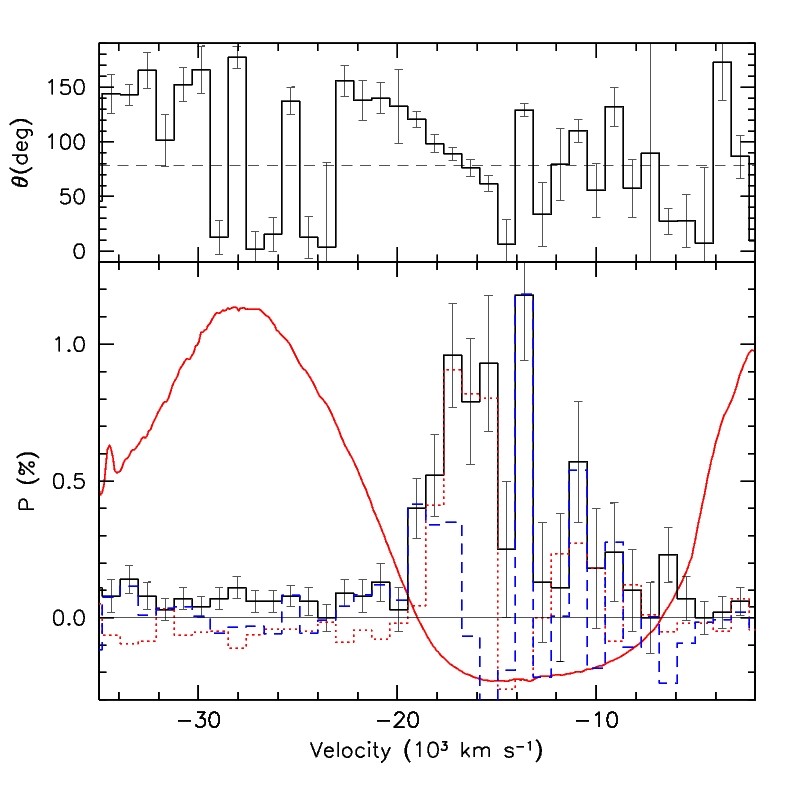}
\caption{\label{fig:caprof3}Same as Fig.~\ref{fig:caprof} for day +39.
The horizontal dashed line in the upper panel indicates the dominant
axis deduced on day $-$7.}
\end{figure}

At early times, the Si~II optical depth is high above some parts of
the Thomson scattering photo-disk but low at others and this
asymmetric "shadowing" turns into a net polarization. As the
photosphere shrinks with time, an increasingly higher
velocity/frequency range can be blocked, because the Si abundance
remains high over a large velocity range.  Eventually, for every point
of the photo-disk, there is a region in the atmosphere where the Si~II
line blocks the direct emission from the photosphere and, thus, the
polarization vanishes.  

For the Si~II line, there is another effect, which is equally
important and goes in the same direction.  The Sobolev optical depth
of Si~II is about 30 to 80 at the photosphere (H\"oflich, M\"uller \& Khokhlov \cite{hoeflich93};
H\"oflich \cite{hoeflich95}), but the chemically
asymmetric outer layers become optically thin due to geometrical
dilution. This is not the case for the Ca~II absorption at several
weeks after maximum light, since the outer calcium layers remain
optically thick because the line opacity is larger by several orders
of magnitude.

The strong polarization shown by Si~II and Ca~II, as opposed to the
lack of detection for O~I despite the similar position in velocity
space, associates SN~2006X to SN~2004dt and leads us to conclude that,
as in that case, the distribution of oxygen is roughly spherically
symmetric but is contaminated by intermediate-mass elements extending
into the oxygen-rich region (Wang et al. \cite{wang06b}). In this
respect, we note that an off-center delayed-detonation model
reproduced rather well the high polarization and polarization angle of
Si and Ca and the low polarization of O during the early stages of
SN~2004dt (H\"oflich et al. \cite{hoeflich06}).

\begin{figure*}
\centering
\includegraphics[width=14cm]{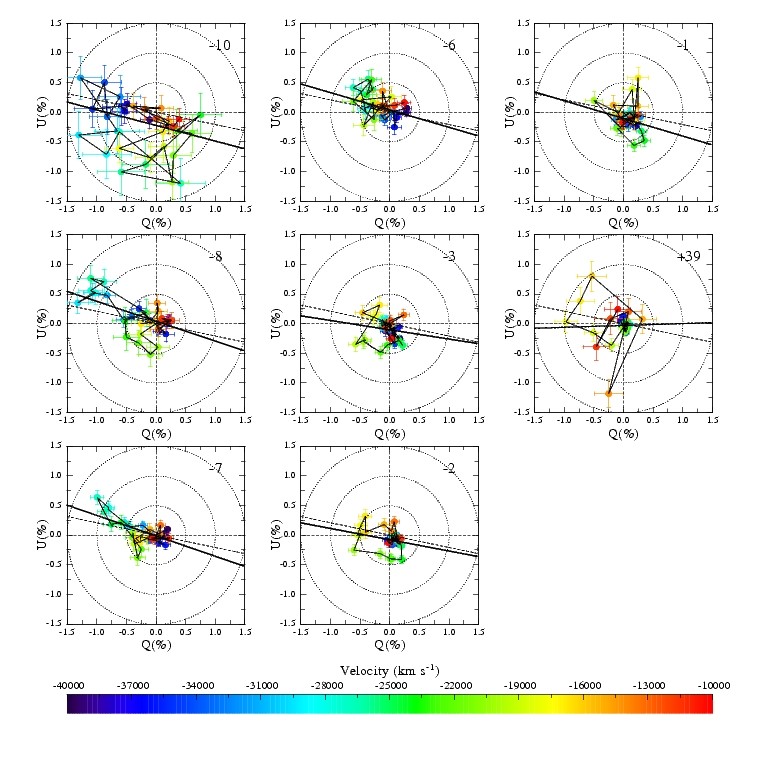}
\caption{\label{fig:dom_ca}$Q$-$U$ diagrams for the Ca~II IR triplet
(see Fig.~\ref{fig:dom_si}). The velocity scale is computed with
respect to the average triplet wavelength (8579.1\AA).}
\end{figure*}

Another feature common to SN~2004dt and SN~2006X is the very high
expansion velocities shown by both SNe. Nevertheless, other properties
make SN~2006X more closely resemble SN~2001el (Wang et
al. \cite{wang03}).  Both SN~2006X and SN~2001el share a larger
intensity of the Ca~II HVF and its distinct polarization parameters
with respect to the photospheric component. At the time of SN~2001el,
the recognition of HVFs was only emerging.  Nevertheless, Wang et al.
(\cite{wang03}) commented that if one of the explanations discussed by
them, namely a torus-like structure, were right, occurrences of strong
HVFs should be strongly polarized (a low number of large clumps
would have a similar effect. See also Kasen et al. \cite{kasen} and
Tanaka et al. \cite{tanaka} for the effects of toroidal distributions
of material).  SN~2004dt has shown that the opposite is not (always)
true: this supernova's Ca II HVF was highly polarized whereas the line
strength was lower than in both SN~2001el and SN~2006X.  In SN~2006X,
the Ca~II IR HVF is strong in both equivalent width and polarization,
thereby matching SN~2001el.  But, unlike SN~2001el, the polarization
angle of the HVF is not very different from the one of the dominant
axis.

\begin{figure}
\centering
\includegraphics[width=8cm]{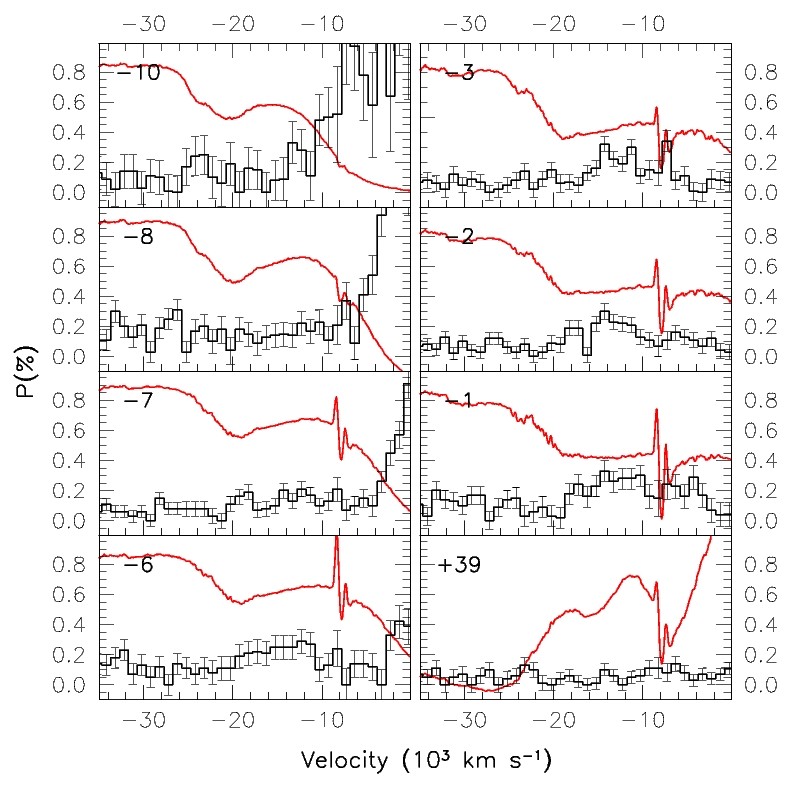}
\caption{\label{fig:oevol}Polarimetric evolution of the O~I 7774\AA\/
line profile. The smooth solid curve traces the unbinned flux
spectrum, arbitrarily scaled for presentation. The broad polarized
feature on the right edge is the blue wing of the Ca~II near-IR
 triplet.}
\end{figure}

\subsection{The near IR Ca~II triplet and its HVF}

One of the most interesting aspects shown by the polarization data of
SN~2006X is the behavior of the Ca~II near-IR triplet and the distinct
polarization properties of the HV component with respect to the
photospheric component (see Figs.~\ref{fig:caprof}, \ref{fig:caprof2}
and \ref{fig:caprof3}). In general, we believe this is consistent with
the Ca~II line-forming regions discussed by Kasen et
al. (\cite{kasen}) and Gerardy et al. (\cite{gerardy}). If the HV
component is formed in a region that asymmetrically shadows the
Thomson scattering photosphere, it will lead to an incomplete
cancellation of the photospheric polarized light. In turn, this can
explain the loop on the $Q$-$U$ plane displayed by the Ca~II HV
component in Fig.~\ref{fig:dom_ca} (Kasen et al. \cite{kasen}; Wang \&
Wheeler \cite{WW08}). Moreover, blending the more polarized HV and
less polarized photospheric components (as is the case for SN~2006X)
also accounts for the observed blue-shift of the polarization peak
with respect to the line center in the flux spectrum.

Clearly, the time evolution shown by the two components indicates a
scenario which is more complicated than that derived for SN~2001el,
for which the Ca~II photospheric component did not show a significant
polarization (Wang et al. \cite{wang03}; Kasen et al. \cite{kasen}).
Also at variance with the case of SN~2001el, SN~2006X shows a
comparable polarization for the Si~II 6355\AA\/ line, and it displays
a loop in the $Q$-$U$ plane prior to maximum light (see
Fig.~\ref{fig:dom_si}), indicating that this feature is formed within
an odd shaped line-forming region, at least in the early epochs.
Moreover, the Si~II and Ca~II loops tend to lie in the $-Q$,+$U$
quadrant of the $Q$-$U$ plane (Figs.~\ref{fig:dom_si},
\ref{fig:dom_ca}). 

This might imply that there is an extended protrusion of peculiar
shape, exterior to the photosphere and photospheric line-forming
regions, that gives rise to the Ca~II HV component and, to a lesser
extent, some of the Si~II feature. It is rather intriguing that the HV
Ca~II and the photospheric Si~II features appear to share the same
dominant axis, at least in the earliest epochs (compare
Fig.~\ref{fig:siprof} to Fig.~\ref{fig:caprof}).

Even though the Ca~II HVFs are a common feature in the early,
pre-maximum spectra of Type Ia SNe (Mazzali et al. \cite{mazzali05a}),
their origin is still debated. The most favored scenario is one where
these features arise within circumstellar material that is overrun by
the rapidly expanding outermost layers of the SN ejecta (Gerardy et
al. \cite{gerardy}). An additional phenomenon is the Ca (and possibly
Si) abundance enhancement in the outer regions of the explosion
(Mazzali et al. \cite{mazzali05a}), which however requires also
density enhancements (i.e. a 3D structure in the explosion) in order
to account for the strength of the HVF (Mazzali et
al. \cite{mazzali05b}). Quimby et al. (\cite{quimby06}) point out that
the CSM interaction picture is supported by the velocity cutoff seen
in the Si~II 6355\AA\/ blue wing profile. In the case of SN~2006X, our
data show that the HVF is blended with the photospheric feature
indicating, in the scenario proposed by Gerardy et
al. (\cite{gerardy}), a high mass for the shell produced by the
interaction (for SN~2003du Gerardy et al. estimated a mass of
$\approx$10$^{-2}$ M$_\odot$; see also Mazzali et
al. \cite{mazzali05a} and Tanaka et al. \cite{tanaka}). This blending
would also be consistent with the Si~II velocity cutoff and the
velocity shift between the HVF and the photospheric component (Gerardy
et al. \cite{gerardy}; Quimby et al. \cite{quimby}) which, on day
$-$10, is about 9,000 km s$^{-1}$ (Fig.~\ref{fig:casi}).

We notice that this value is higher than the $\sim$7000 km s$^{-1}$
seen in 2002bo (Mazzali et al. \cite{mazzali05a}) and 2004du (Gerardy
et al. \cite{gerardy}).  Shell masses apparently vary in a wide range
(e.g.  $\approx$5$\times$10$^{-3}$ M$_\odot$ for SN~2005cg, Quimby et
al. \cite{quimby}, to 5$\times$10$^{-2}$ M$_\odot$ for SN~2002bo,
Gerardy et al. \cite{gerardy}) and, consequently, the cutoff velocity
is also expected to do so.  Whether this is related to the extremely
high velocities recorded for SN~2006X or is rather due to some other
effect, remains to be clarified.

Normally the HVF are well visible in the very early stages of the SN
evolution and disappear around maximum light (Mazzali et
al. \cite{mazzali05a}). 
The spectropolarimetry shows a clear component at about $-$25,000 km
s$^{-1}$ on day $-$1, to be compared with the $-$16,300 km s$^{-1}$
deduced from the minimum of the Si~II 6355\AA\/ absorption
trough. From this we deduce that the HVF component is still present
around maximum light, and its blue-shift has decreased by about 2,000
km s$^{-1}$ from day $-$8. We notice that this variation is not very
different from the one shown by the expansion velocity in the same
time interval ($\sim$2,500 km s$^{-1}$, see Fig.~\ref{fig:vel}).

If the HV Ca feature is indeed produced by the interaction
between the SN ejecta and the surrounding ``non-processed''
material, then the polarization observed across the HVF implies that
this material is distributed in an asymmetric way.

\begin{figure}
\centering
\includegraphics[width=8cm]{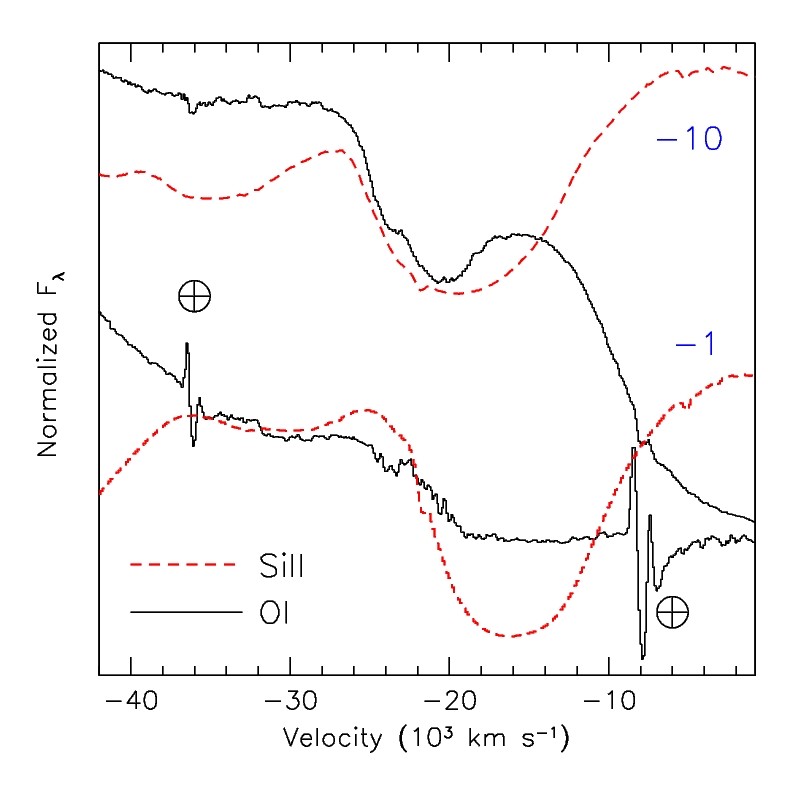}
\caption{\label{fig:osi}O~I 7774\AA\/ and Si~II 6355\AA\/ line profiles 
on days $-$10 (top) and $-$1 (bottom).}
\end{figure}

\subsection{Ca~II re-polarization}

\begin{figure*}
\centering
\includegraphics[width=14cm]{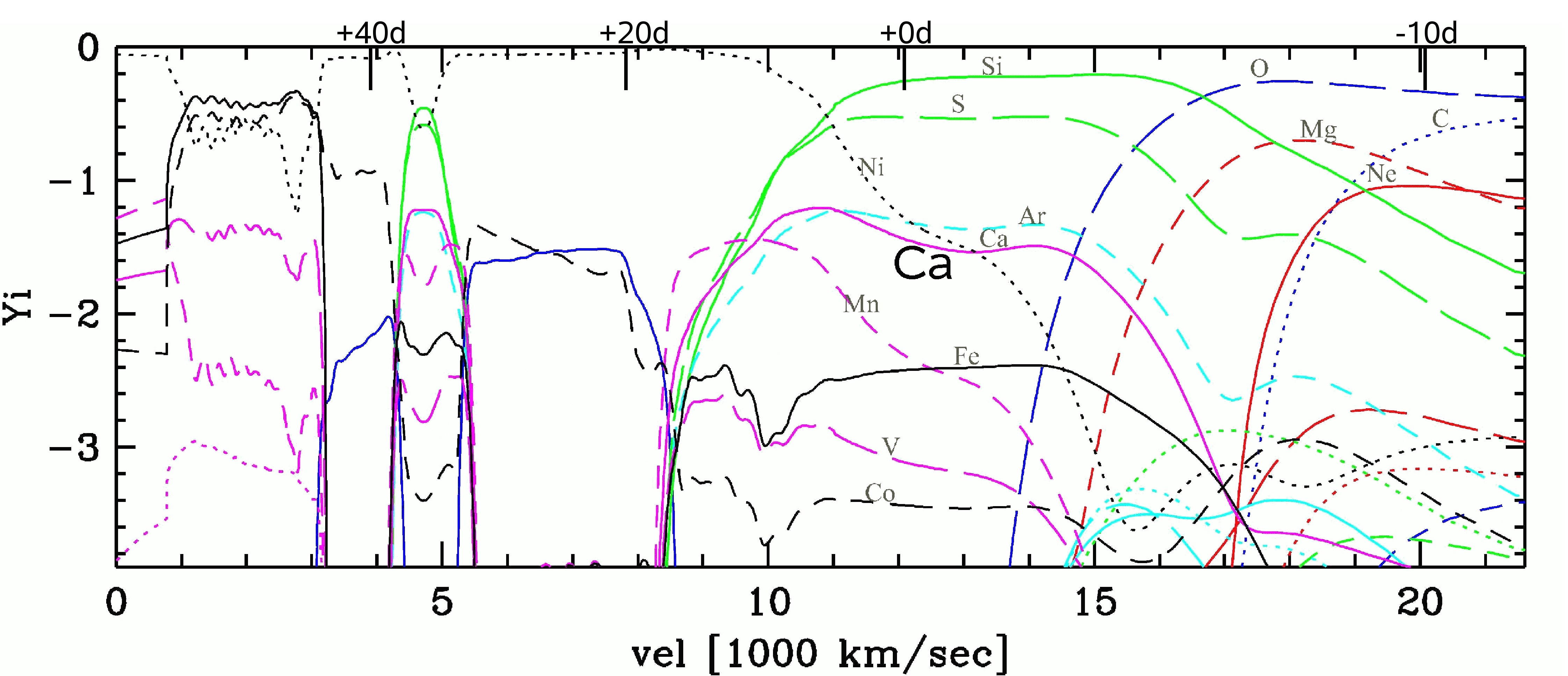}
\caption{\label{fig:DDT} Abundances as a function of expansion velocity
for a delayed detonation model of a $M_V$=$-$19.29, spectroscopically
normal SN~Ia originating from a star with a main sequence mass of 5
$M_\odot $. The deflagration/detonation transition density is
2.5$\times$10$^7$ g cm$^{-3}$ (H\"oflich et al. \cite{hoeflich02}).
For all but very early times, Ca and Si are singly ionized. Tickmarks
on the upper abscissa mark the velocities of the Thomson-scattering
photosphere at various times relative to maximum light. Within this
series of delayed-detonation models for spectroscopically normal SNe
Ia, $M_V$ is expected to range between $-$18.98 and $-$19.35, and
expansion velocities vary by about 15\%.}
\end{figure*}

The most striking event in the evolution of SN~2006X is the
polarization observed on day +39, peaking at about $-$16,000 km
s$^{-1}$ (Fig.~\ref{fig:caprof3}). The absorption profile is rather
broad for this relatively late epoch: it is broader than on day $-$1
(Fig.~\ref{fig:caevol}) and in SN~2002bo at a comparable epoch
(Fig.~\ref{fig:ca}). The difference with SN~2002bo is produced by the
strong component at $-$16,000 km s$^{-1}$ that gives rise to a marked
line profile asymmetry.

The expansion velocity deduced from the Si~II 6355\AA\/ for day +30 is
about 12,000 km s$^{-1}$ (Wang et al. \cite{wangx}), while the minimum
of the absorption trough in our last spectrum is $\sim$13,000 km
s$^{-1}$. A lower value ($\sim$9,500 km s$^{-1}$) is deduced
by interpolating the Fe~II 4555\AA\/ line velocities reported by Wang et
al. (\cite{wangx}) to day +40.  Strong lines are formed well above the
photosphere and, if the abundance of a specific element declines, then the
expansion velocity does not follow the receding photosphere but
levels off in velocity space as in the case of Si II (e.g. H\"oflich
\& Khokhlov \cite{hk96}, Benetti et al. \cite{benetti04}). 
The fact that on the same late epoch there is no trace of Si~II
6355\AA\/ line polarization (Fig.~\ref{fig:sievol}) indicates that
there must be a substantial difference in optical depth and geometry 
between the regions where calcium and silicon lines are generated in 
this phase.

The disappearance of the polarization in the Si features can be
understood in the framework of off-center DDT models. Early on, the Si
lines are formed in a region where, for a given radius, the Si
abundance is direction dependent. This causes an 'asymmetric' covering
of the photosphere (see Fig.~2 in H\"oflich \cite{hoeflich06}) leading
to a net polarization.  As time goes by, the photosphere and the Si
line forming region recede well within the Si-rich region, so that
there is no selective 'covering' of the photosphere and hence no Si
polarization.  For this reason the Si polarization disappearance is a
general phenomenon, expected to take place in any scenario/model with
an extended, aspherical region of incomplete burning
(Si/S). Therefore, it is not limited to off-center DDT or DD models,
provided the SN undergoes a detonation phase with O-burning.

\begin{figure}
\centering
\includegraphics[width=8cm]{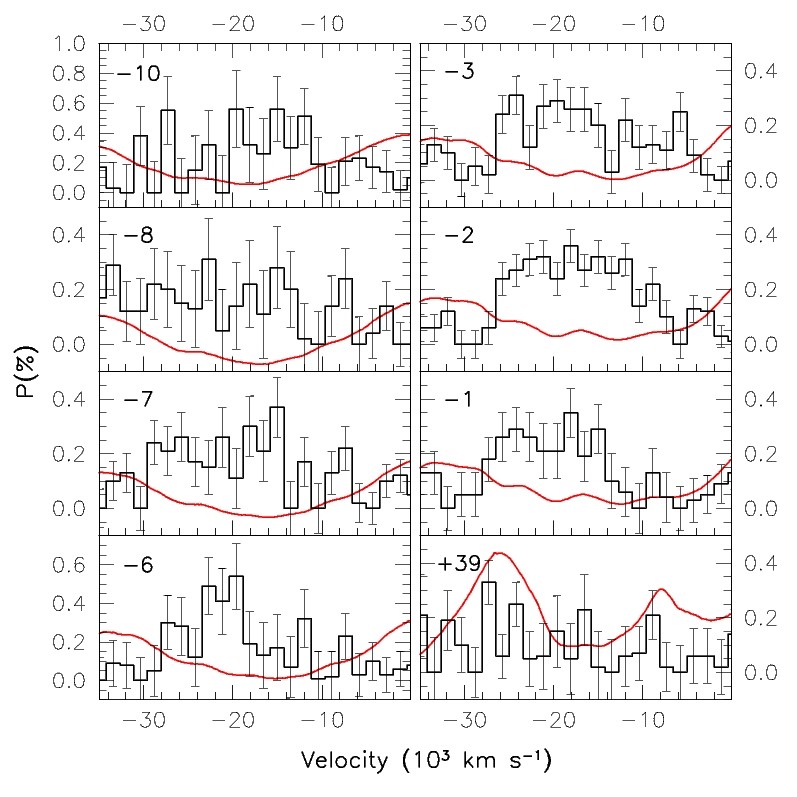}
\caption{\label{fig:si2evol}Polarimetric evolution of the Si~II 5051\AA\/
line profile. The smooth solid curve traces the unbinned flux
spectrum, arbitrarily scaled for presentation.}
\end{figure}

At our last epoch, about 40 days past maximum light, the photosphere
has receded below 5000 km s$^{-1}$, well into the $^{56}$Ni-rich
region (Fig.~\ref{fig:DDT}), where Ca is a product of the explosive
nucleosynthesis and its abundance is about 1 to 2\%, i.e. enhanced by
a factor of $\approx$10$^4$ with respect to its solar value. For a
spectroscopically normal SN~Ia (Hoeflich et al. \cite{hoeflich02}), a
Ca-rich region of almost constant abundance is predicted between 9,000
to 15,000 km s$^{-1}$ (Fig.~\ref{fig:DDT})\footnote{The Ca-rich layer
seen in Fig.~\ref{fig:DDT} at about 5000 km s$^{-1}$ is an artifact of
spherically symmetric delayed-detonation models and is related to the
fact that the initiation of the delayed-detonation is located in a
shell (Khokhlov \cite{khokhlov95}; H\"oflich \cite{hoeflich95}).}.
The abundance of Ca declines logarithmically and smoothly outward,
converging to a solar value between 17,500 to 19,000 km s$^{-1}$. This
is particularly interesting, because the outer edge (placed at about
15,000-17,000 km s$^{-1}$; see Fig.~\ref{fig:DDT}) coincides with the
velocity at which we see strong polarization several weeks after
maximum (Fig.~\ref{fig:caprof3}), when the scattering-dominated
photosphere is receding through the $^{56}$Ni region.

\begin{figure}
\centering
\includegraphics[width=8cm]{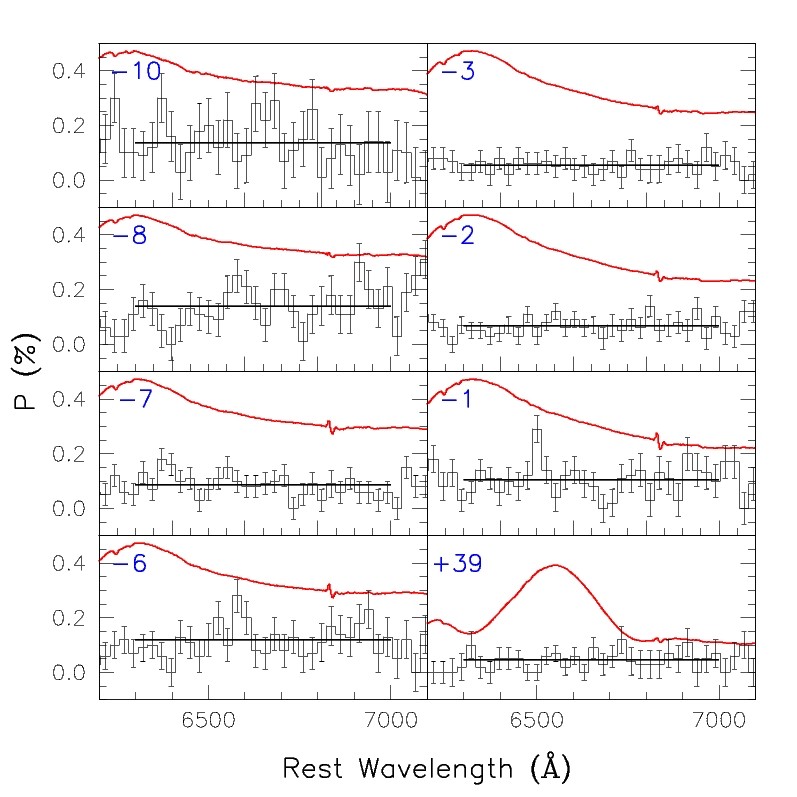}
\caption{\label{fig:contevol}Polarimetric evolution of the 
6200-7000\AA\/ continuum. The smooth solid curve traces the unbinned
flux spectrum, arbitrarily scaled for presentation. The horizontal
line is placed at the average polarization level within the wavelength
range 6300-7000\AA.}
\end{figure}

Note that theoretical spectra do not show a significant polarization
in the photospheric component of Ca II during the early times when the
photosphere recedes from 17,000 to 12,000 km s$^{-1}$ (H\"oflich et
al. \cite{hoeflich06}) because of the rather smooth decline in
abundance and the huge cross section of the Ca-IR triplet which causes
the line to be optically thick ($\tau\approx 1$), even for abundances
close to solar (Fig.~\ref{fig:DDT}).  The Ca~II lines cover up the
entire photosphere, leading to a low polarization of the
photospheric component similar to that displayed by oxygen, although
the asymmetries in Ca caused by off-center delayed detonation (see the
discussion in Sect.~\ref{sec:o}) are similar in size to those of the
Si-rich layers (Fesen et al. \cite{fesen}). Similarly to the behavior
of Si, the polarization is low when the photosphere is well within
the Ca-rich layers because of the large extension of Ca in velocity
space ($\approx$8000 to 17000 km s$^{-1}$, see Fig.~\ref{fig:DDT}; see
also Branch et al. \cite{branch06}) and because almost all of the
inner, Thomson scattering dominated atmosphere is covered up by Ca at
a range of expansion velocities.

Several weeks after maximum cooling and geometric dilution cause the
optical depth to decrease by more than an order of magnitude at the
outer boundary of the Ca-rich region. By that time, the photosphere
has receded within the inner edge of the Ca layer at 8000-9000 km
s$^{-1}$.  The result is a partial cover up of the photosphere in the
line wings that generate a polarization comparable to the one shown
by Si, because of the similar size of asymmetry of the Si and Ca
layers.  The large extension of Ca in velocity space ($\Delta$v=10,000
km s$^{-1}$) is consistent with its observed broad absorption. It is
expected that a large fraction of the inner Thomson scattering region
(v$_{ph}\approx$ 4000 km s$^{-1}$) is covered-up by Ca at most
wavelengths. In the line wings where the Sobolev optical depth of the
line drops to a few, the blocking is less, resulting in high
polarization. The fact that we do not see a statistically significant
polarization at the inner Ca boundary (v$\lesssim$10,000 km s$^{-1}$;
see Fig.~\ref{fig:caprof3}) is a probable sign of mixing at the inner
edge of the Ca-rich layers (8000 km s$^{-1}$, see Fig.~\ref{fig:DDT}).

\begin{figure}
\centering
\includegraphics[width=8cm]{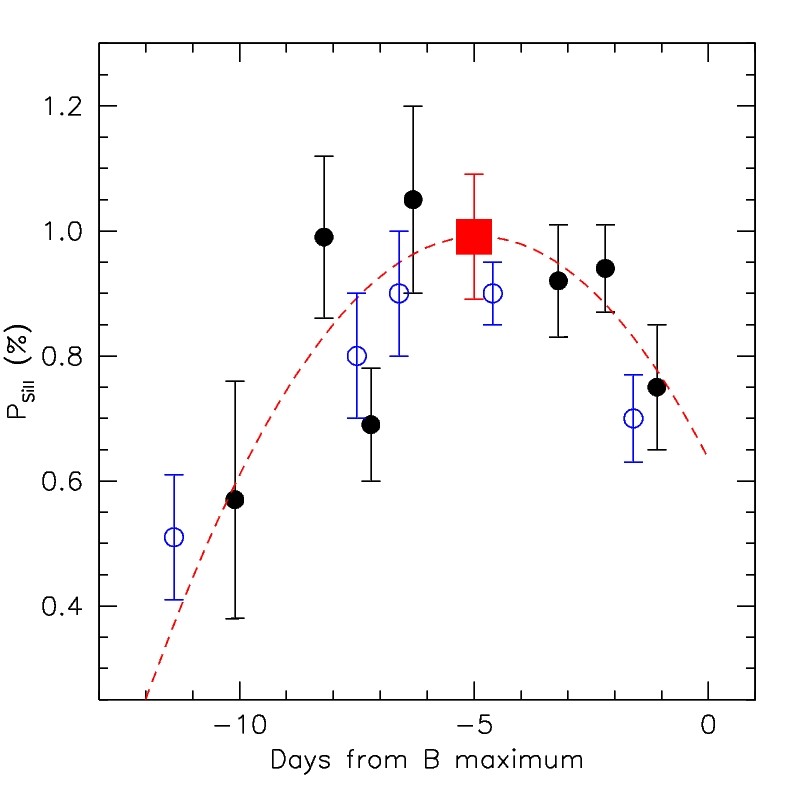}
\caption{\label{fig:sipol}Linear polarization of Si~II 6355\AA\/ 
as a function of epoch. The dashed line is a second order polynomial
fitting to SN~2006X data and the filled square dot is the polarization
interpolated for day $-$5 (see Sect.~\ref{sec:gen}). The empty circles
are the data for SN~2002bo (Wang et al. \cite{wang07}).}
\end{figure}

Unfortunately, a single observation does not provide information
separately for the size of the asymmetry, the orientation of the
observer and mixing processes which are expected to ``soften'' the
steep gradients in layers with rapid changes of abundances.  It is the
time dependence, which is direction dependent for a given supernova,
that tells us how fast the photosphere passes a given chemical
``boundary layer''.  This is why it will be very important, in the
future, to follow Type Ia SNe with multi-epoch spectropolarimetry at
epochs later than a month.  This will allow us to understand whether
the re-polarization is a transient phenomenon (as it should be) and
tell us something about mixing. In particular, the mixing component
should introduce a change of the dominant axis with time on short
time-scales, whereas large scale asymmetries should cause slow
variations. Time coverage would thus allow us to separate the
different components.  In this respect, for the case of SN~2006X, the
data are inconclusive. While a dominant axis is present before maximum
light (Fig.~\ref{fig:caprof},
\ref{fig:caprof2}), it is rather unclear at the last epoch. There
might be an indication that the $-$16,000 km s$^{-1}$ component has a
polarization angle similar to the one displayed by the feature at a
similar velocity just before maximum light (Fig.~\ref{fig:caprof3}),
but we do not attach too much confidence to this fact.

Because basically all observations have shown SNe Ia to become
unpolarized within a week after $B$ maximum light, observational
series have routinely been discontinued shortly after peak brightness;
but 2 out of 2 SNe Ia observed at considerably later phases (SN~2006X
and SN~2001el) now show convincingly the tomographic power given by
the combination of spectroscopy and polarimetry. The case of SN~2006X
demonstrates that spectropolarimetry enables one to probe regions which are
dominated by the early phases of the deflagration burning and the
thermonuclear runaway.  Late time polarization, both in size and
timing, will depend on the viewing angle and on a variety of other
aspects. Among these are the location at which the deflagration front
turns into a detonation and the amount of mixing produced by
instabilities during the deflagration phase (plumes), which may wash
out abundance gradients by different amounts (Fesen et
al. \cite{fesen}, Hoeflich et al. \cite{hoeflich02}).

While SN~2001el did not show polarization at late epochs, SN~2006X
certainly did. This may be due to the quality and timing of
observations, orientation effects or may hint at an intrinsic
diversity, thus providing a glimpse of what we might learn in the
future by combining extensive spectropolarimetric coverage with
systematic theoretical studies.

\begin{figure}
\centering
\includegraphics[width=8cm]{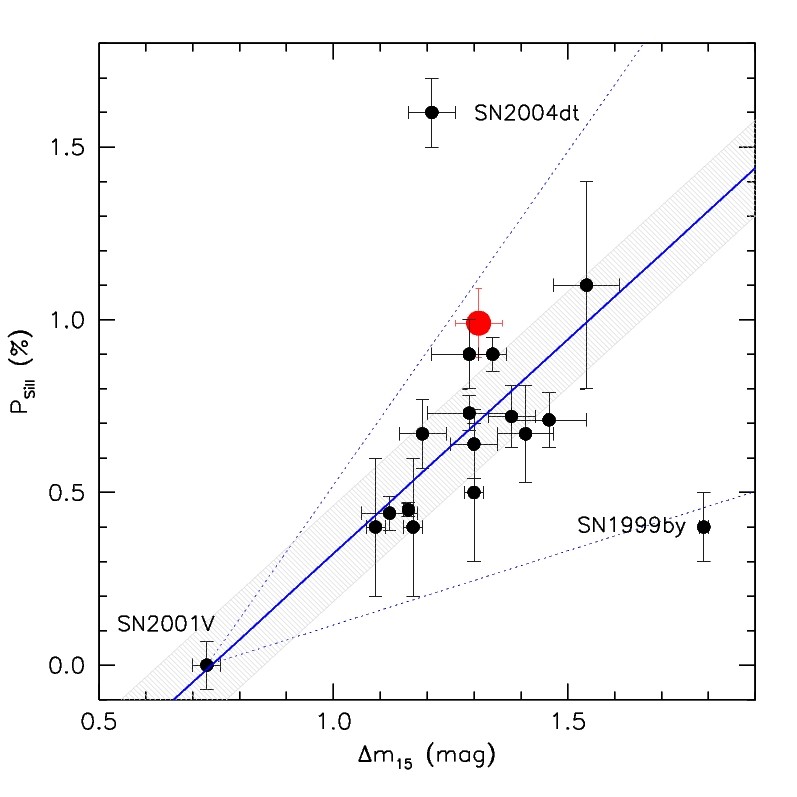}
\caption{\label{fig:poldm}Degree of polarization across 
the Si~II 6355\AA\/ line as a function of light curve decline
rate. Data are from Wang et al. (\cite{wang07}). SN~2006X is marked by
the large dot. The solid line is a best fit to all data, with the
exception of SN~1999by and SN~2004dt. The dotted lines trace the
1-$\sigma$ level of the intrinsic polarization distribution generated
by the Monte-Carlo simulation discussed by Wang et
al. (\cite{wang07}). The shaded area indicates the rms deviation of
the data points from the best fit relation.}
\end{figure}

\begin{acknowledgements}
This paper is based on observations made with ESO Telescopes at
Paranal Observatory under program IDs 076.D-0177(A) and 076.D-0178(A).
The authors are grateful to ESO-Paranal staff for the support given
during the service mode. This work is partially based on NSF grants
AST 04-06740,07-03902 \& 07-08855 to P.A.H., AST-0707769 to J.C.W.
and AST-0708873 to L.W.. F.P. wishes to thank Dr  P. Danielson for her
kind help.
\end{acknowledgements}

\Online

\appendix

\section{\label{sec:app} Additional figures}

\begin{figure*}
\centering
\includegraphics[width=10cm]{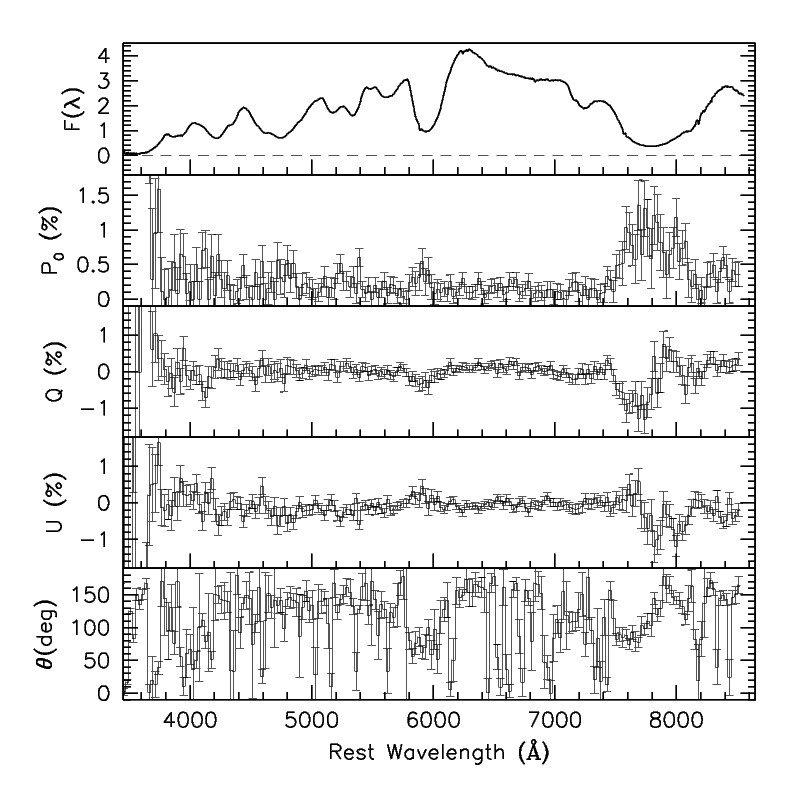}
\caption{\label{fig:pol-10}Spectropolarimetric data for SN~2006X on day $-$10.
From top to bottom: total flux spectrum (in 10$^{-15}$ erg s$^{-1}$
cm$^{-2}$ \AA$^{-1}$), ISP corrected polarization degree, the Stokes
parameters $Q$ and $U$ and polarization position angle on the plane of
the sky.}
\end{figure*}

\begin{figure*}
\centering
\includegraphics[width=10cm]{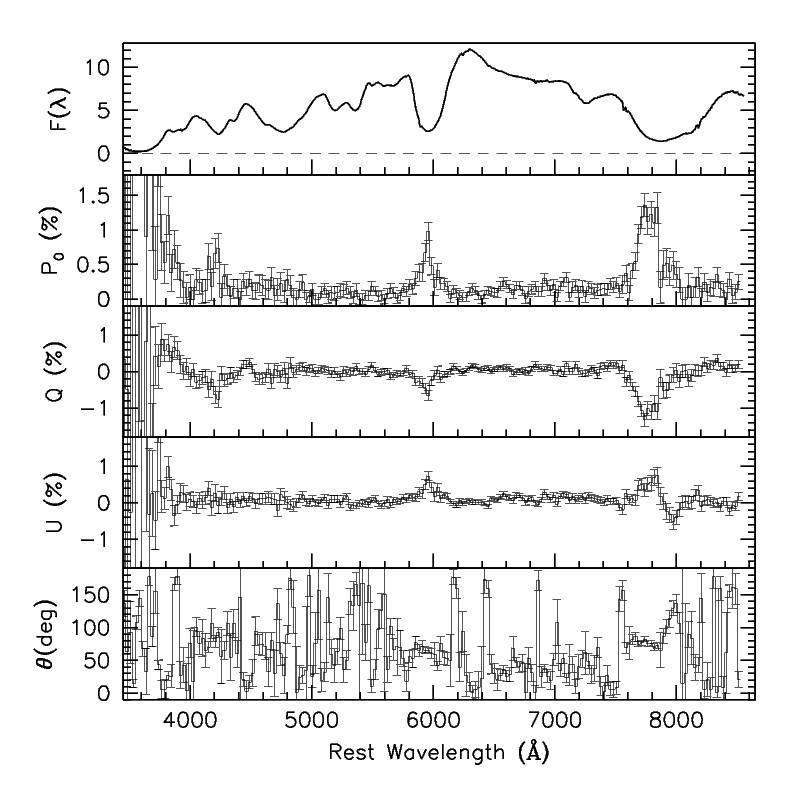}
\caption{\label{fig:pol-8}Same as Fig.~\ref{fig:pol-10} for day $-$8.}
\end{figure*}

\begin{figure*}
\centering
\includegraphics[width=10cm]{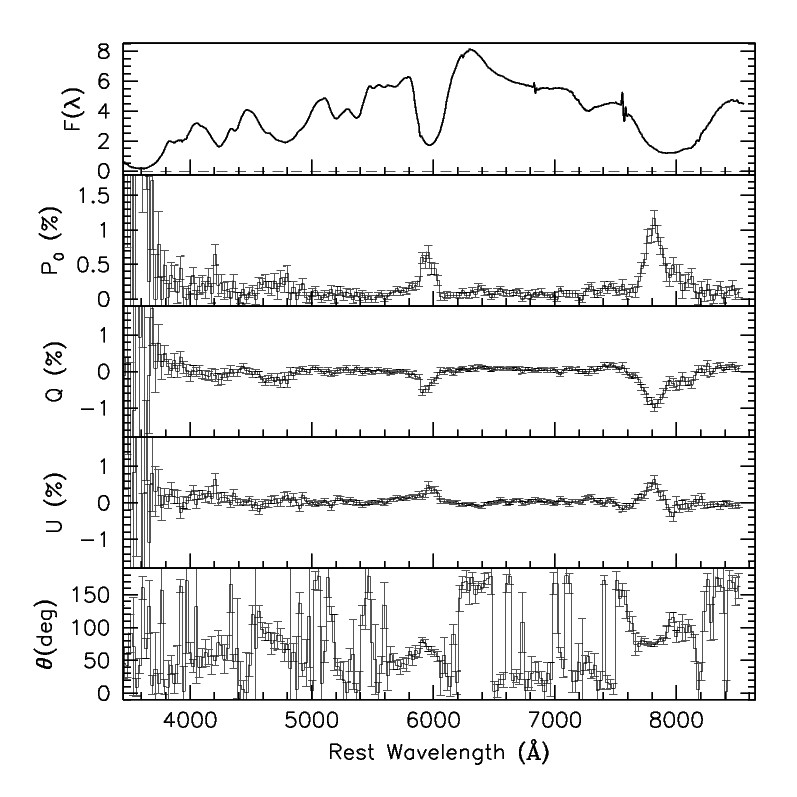}
\caption{\label{fig:pol-7}Same as Fig.~\ref{fig:pol-10} for day $-$7.}
\end{figure*}

\begin{figure*}
\centering
\includegraphics[width=10cm]{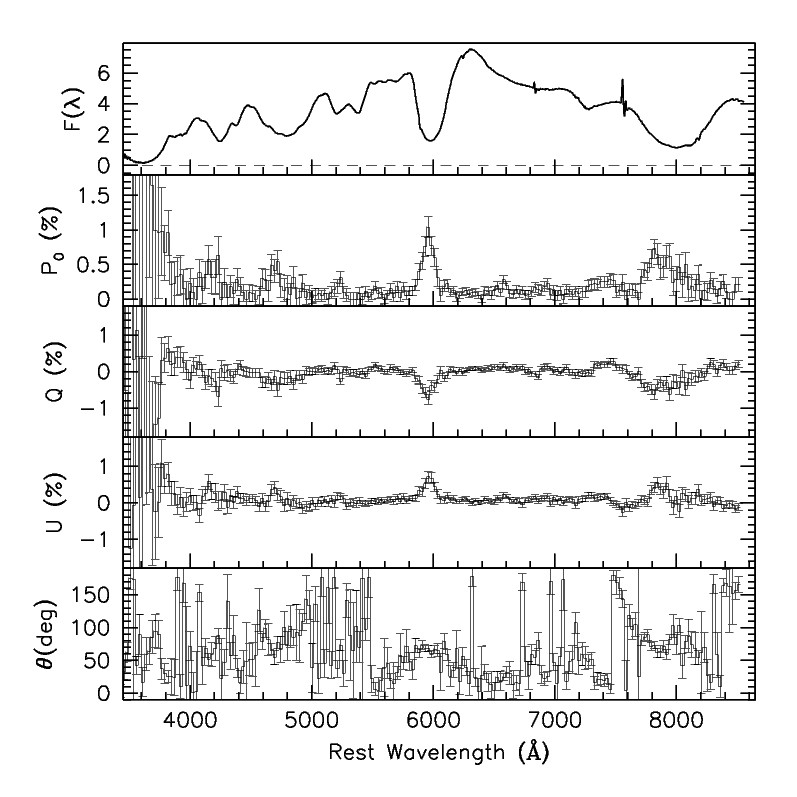}
\caption{\label{fig:pol-6}Same as Fig.~\ref{fig:pol-10} for day $-$6.}
\end{figure*}

\begin{figure*}
\centering
\includegraphics[width=10cm]{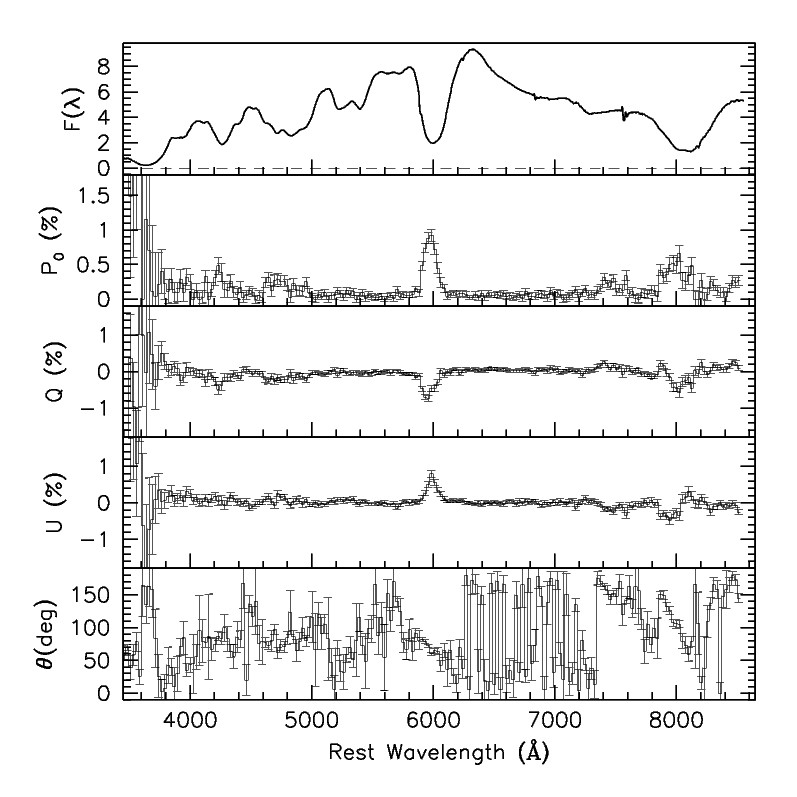}
\caption{\label{fig:pol-3}Same as Fig.~\ref{fig:pol-10} for day $-$3.}
\end{figure*}

\begin{figure*}
\centering
\includegraphics[width=10cm]{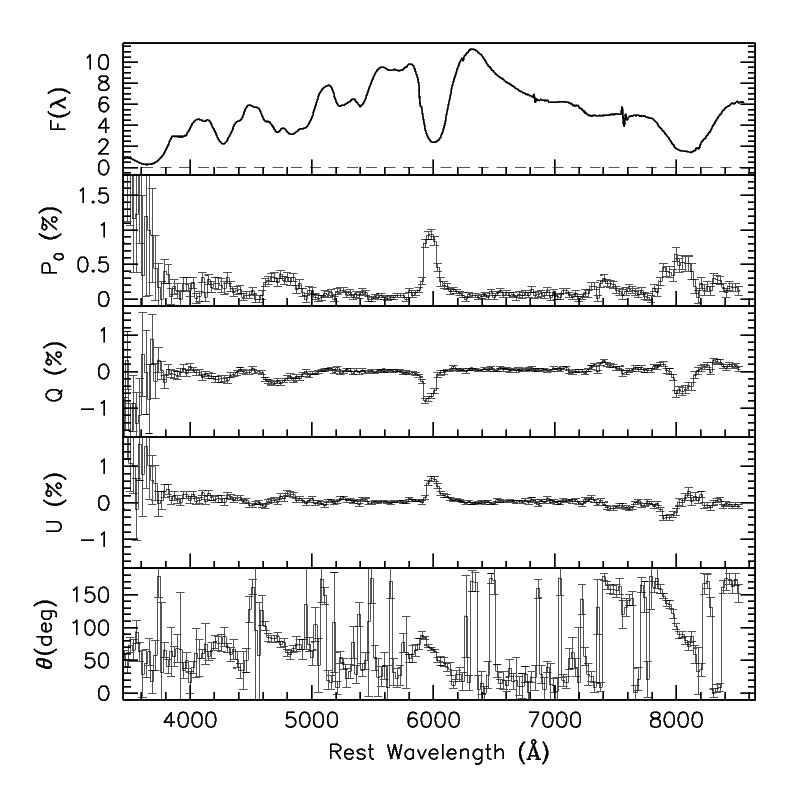}
\caption{\label{fig:pol-2}Same as Fig.~\ref{fig:pol-10} for day $-$2.}
\end{figure*}

\begin{figure*}
\centering
\includegraphics[width=10cm]{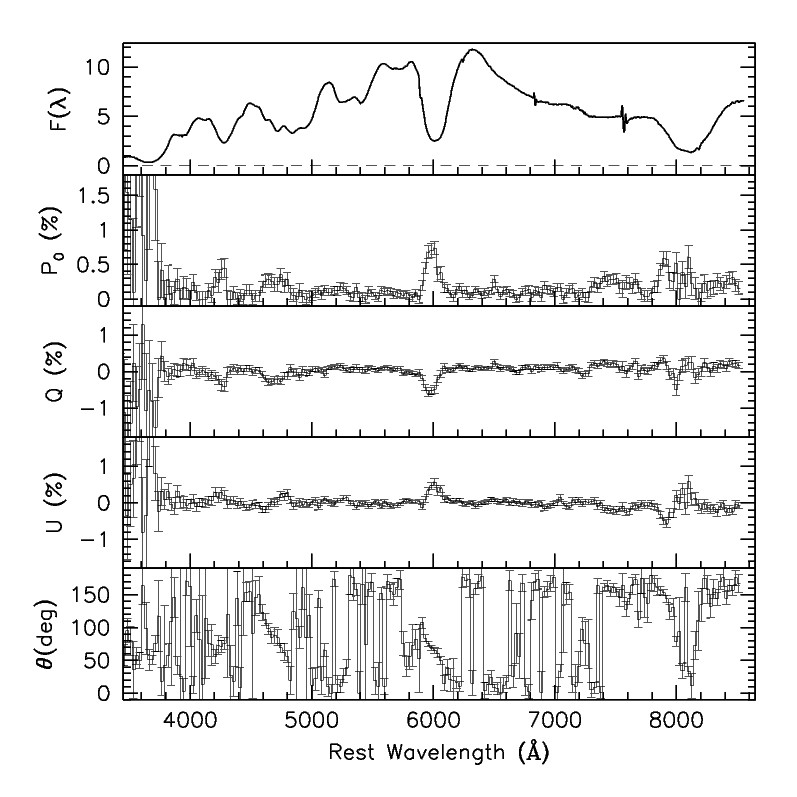}
\caption{\label{fig:pol-1}Same as Fig.~\ref{fig:pol-10} for day $-$1.}
\end{figure*}

\begin{figure*}
\centering
\includegraphics[width=10cm]{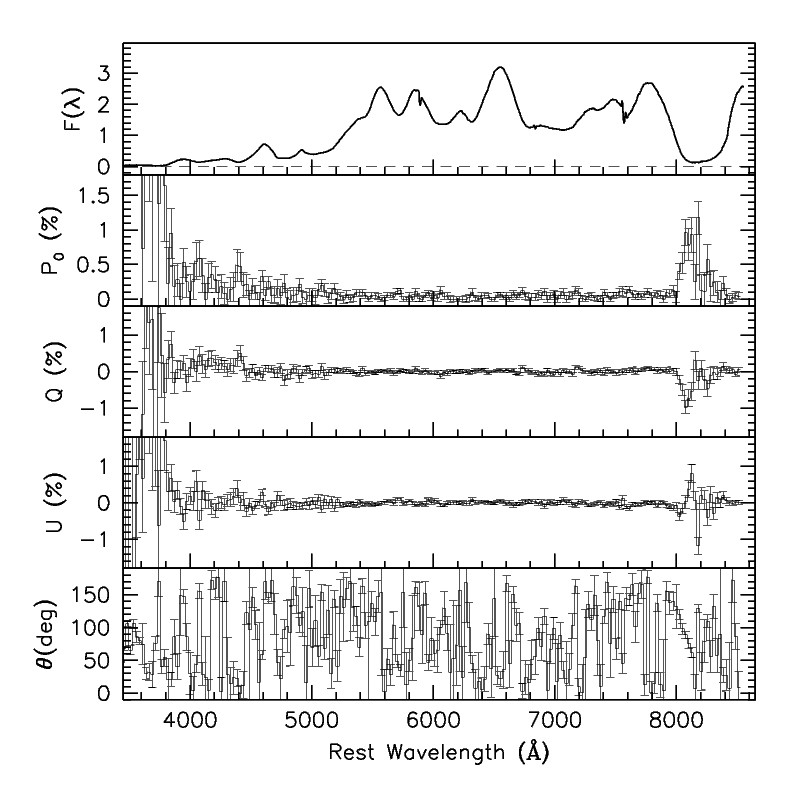}
\caption{\label{fig:pol+39}Same as Fig.~\ref{fig:pol-10} for day +39.}
\end{figure*}

\begin{figure*}
\centering
\includegraphics[width=10cm]{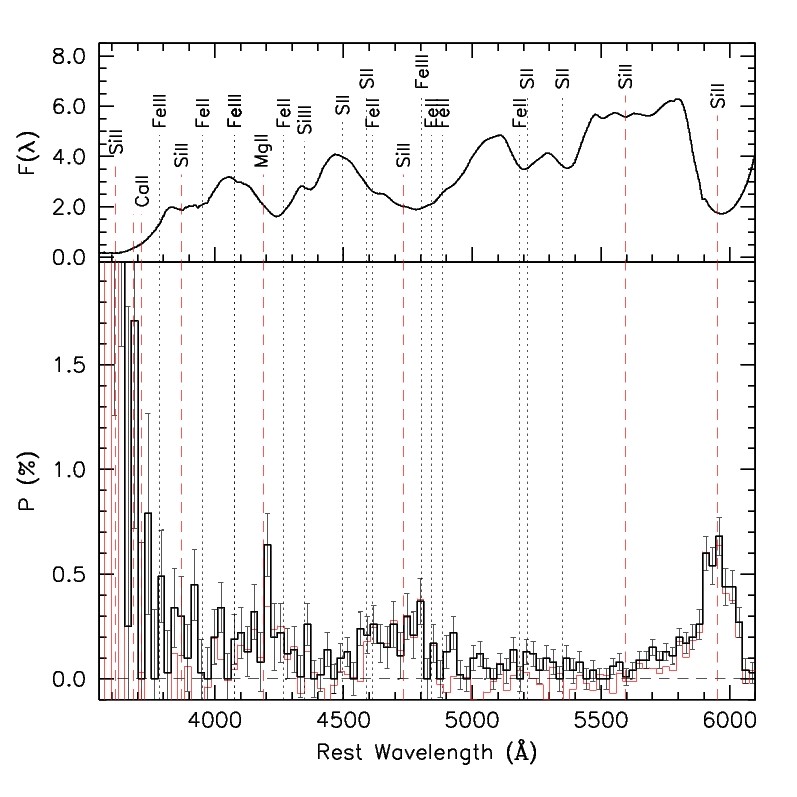}
\caption{\label{fig:polm7}Spectropolarimetry of SN~2006X on day $-$7.
Main line identifications are given in the upper panel. The red
histogram traces the polarization along the dominant axys.  Vertical
dashed lines correspond to an expansion velocity of 19,000 km
s$^{-1}$, while the dotted lines correspond to 14,000 km s$^{-1}$.}
\end{figure*}

\begin{figure*}
\centering
\includegraphics[width=10cm]{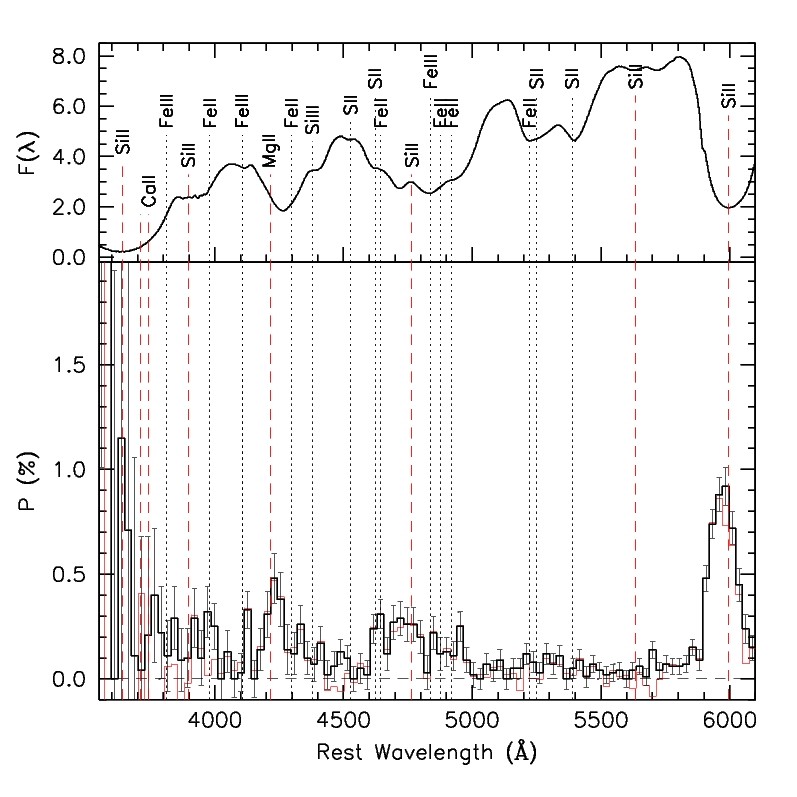}
\caption{\label{fig:polm3} Same as Fig.~\ref{fig:polm7} for day $-$3. 
Vertical dashed lines correspond to a expansion velocity of 17,000
km s$^{-1}$, while the dotted lines correspond to 12,000 km s$^{-1}$.}
\end{figure*}

\begin{figure*}
\centering
\includegraphics[width=10cm]{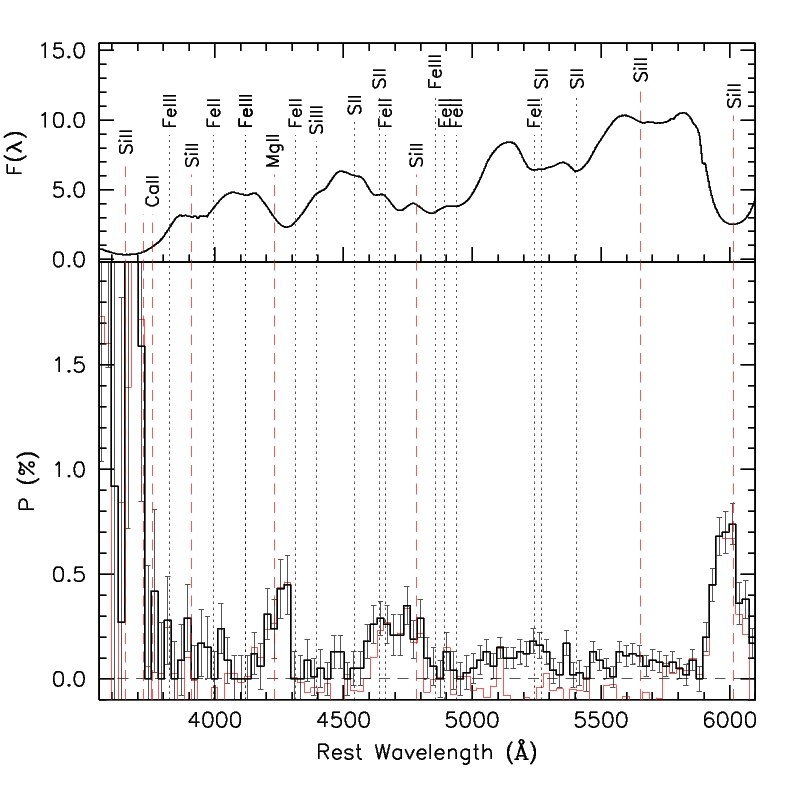}
\caption{\label{fig:polm1} Same as Fig.~\ref{fig:polm7} for day $-$1. 
Vertical dashed lines correspond to an expansion velocity of 16,000
km s$^{-1}$, while the dotted lines correspond to 11,000 km s$^{-1}$.}
\end{figure*}

\begin{figure*}
\centering
\includegraphics[width=10cm]{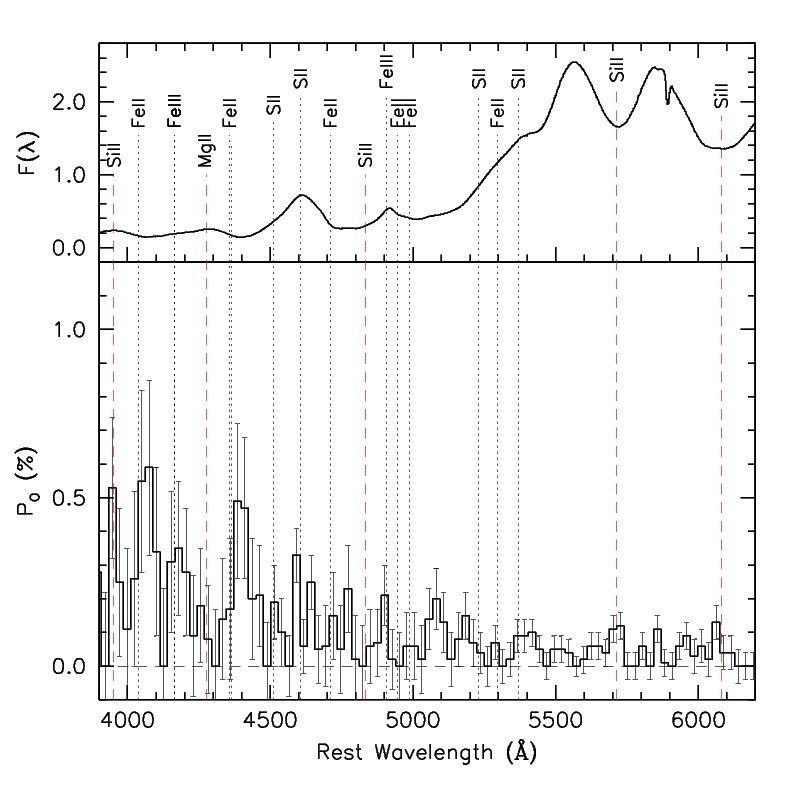}
\caption{\label{fig:polp39} Same as Fig.~\ref{fig:polm7} for day +39. 
Vertical dotted lines correspond to a expansion velocity of 13,000
km s$^{-1}$.}
\end{figure*}

\end{document}